\pdfoutput=1

\documentclass[aps,reprint,superscriptaddress,floatfix]{revtex4-1}           

\renewcommand{\cdot}{} 

\usepackage{graphicx}
\usepackage{amsmath}

\usepackage[caption=false]{subfig}

\begin{document}


\title{The impact of classical control electronics on qubit fidelity}


\author{J.P.G. van Dijk}
\email[]{J.P.G.vanDijk@tudelft.nl}
\affiliation{QuTech, Delft University of Technology, P.O. Box 5046, 2600 GA Delft, The Netherlands}
\affiliation{Kavli Institute of Nanoscience, P.O. Box 5046, 2600 GA Delft, The Netherlands}
\author{E. Kawakami}
\affiliation{Okinawa Institute of Science and Technology, Okinawa 904-0412, Japan}
\author{R.N. Schouten}
\affiliation{QuTech, Delft University of Technology, P.O. Box 5046, 2600 GA Delft, The Netherlands}
\affiliation{Kavli Institute of Nanoscience, P.O. Box 5046, 2600 GA Delft, The Netherlands}
\author{M. Veldhorst}
\affiliation{QuTech, Delft University of Technology, P.O. Box 5046, 2600 GA Delft, The Netherlands}
\affiliation{Kavli Institute of Nanoscience, P.O. Box 5046, 2600 GA Delft, The Netherlands}
\author{L.M.K. Vandersypen}
\affiliation{QuTech, Delft University of Technology, P.O. Box 5046, 2600 GA Delft, The Netherlands}
\affiliation{Kavli Institute of Nanoscience, P.O. Box 5046, 2600 GA Delft, The Netherlands}
\affiliation{Intel Corporation, 2501 NW 229th Ave, Hillsboro, OR 97124, USA}
\author{M. Babaie}
\affiliation{QuTech, Delft University of Technology, P.O. Box 5046, 2600 GA Delft, The Netherlands}
\author{E. Charbon}
\affiliation{QuTech, Delft University of Technology, P.O. Box 5046, 2600 GA Delft, The Netherlands}
\affiliation{Kavli Institute of Nanoscience, P.O. Box 5046, 2600 GA Delft, The Netherlands}
\affiliation{Intel Corporation, 2501 NW 229th Ave, Hillsboro, OR 97124, USA}
\affiliation{\'{E}cole Polytechnique F\'{e}d\'{e}rale de Lausanne, Case postale 526, CH-2002 Neuch\^{a}tel, Switzerland}
\author{F. Sebastiano}
\affiliation{QuTech, Delft University of Technology, P.O. Box 5046, 2600 GA Delft, The Netherlands}


\date{\today}

\begin{abstract}
Quantum processors rely on classical electronic controllers to manipulate and read out the quantum state. As the performance of the quantum processor improves, non-idealities in the classical controller can become the performance bottleneck for the whole quantum computer. 
%
%
%
To prevent such limitation, this paper presents a systematic study of the impact of the classical electrical signals on the qubit fidelity. All operations, i.e.~single-qubit rotations, two-qubit gates and read-out, are considered, in the presence of errors in the control electronics, such as static, dynamic, systematic and random errors. Although the presented study could be extended to any qubit technology, it currently focuses on single-electron spin qubits, because of several advantages, such as purely electrical control and long coherence times, and for their potential for large-scale integration.

As a result of this study, detailed electrical specifications for the classical control electronics for a given qubit fidelity can be derived, as demonstrated with specific case studies. 
We also discuss the effect on qubit fidelity of the performance of the general-purpose room-temperature equipment that is typically employed to control the few qubits available today. 
Ultimately, we show that tailor-made electronic controllers can achieve significantly lower power, cost and size, as required to support the scaling up of quantum computers.

\end{abstract}


\maketitle

\section{Introduction}

\begin{figure}
\includegraphics[width=\linewidth]{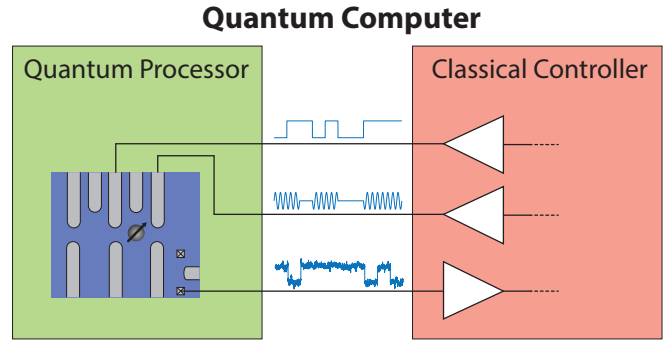}
\caption{\label{fig:quantum_computer} A quantum computer comprises a quantum processor and a classical controller. The classical controller provides the signals for the control of the quantum processor, and processes the signals from the quantum processor to read out the quantum state \cite{patra2017cryocmos}.}
\end{figure}

Quantum computers have the potential to solve problems that are intractable even for the most powerful supercomputers \cite{Montanaro2016}. These problems include the factorization of prime numbers using Shor's algorithm \cite{shor1994algorithms}, the efficient search in large datasets using Grover's algorithm \cite{grover1996fast}, and the simulation of quantum systems as initially proposed by Feynman \cite{feynman1982simulating}.
%
A quantum computer operates by processing the information stored in quantum bits (qubits), which are organized in a quantum processor. Performing operations on the qubits requires a classical electronic controller for manipulating the qubits and reading out their quantum state (Fig.\,\ref{fig:quantum_computer}) \cite{patra2017cryocmos}.
In order not to degrade qubit performance, the classical controller must provide high-accuracy low-noise control signals and the read-out must be very sensitive and quiet to detect the weak signals from the quantum processor without altering the qubit states.

Since state-of-the-art quantum processors comprise only a few qubits ($<$ 20 qubits \cite{IBM17qubits,kelly2015state,monz201114}), the classical controller is currently implemented by general-purpose instruments operating at room temperature. The use of these high-performance instruments results in the fidelity of the quantum operations being limited by the quantum processor \cite{yoneda201799}.
However, as the performance of the quantum processor improves, the classical controller can become performance limiting. Consequently, it is crucial to understand how the controller impacts the performance of the whole quantum computer in order to properly co-design the controller and the quantum processor and to identify potential performance bottlenecks. 

Moreover, the simplest non-trivial algorithms, such as quantum chemistry problems \cite{Wecker2014}, require more than 100 logical qubits. This translates into the need for   thousands or millions of physical qubits, if taking into account the redundancy added by quantum error correction schemes, such as surface codes \cite{Fowler2012}. For such large-scale quantum processors, implementing the classical controller with general-purpose instruments would be impractical and offer limited scalability due to its size and cost.
A more practical and power-efficient approach would be to use tailor-made electronics that can be optimized for this specific application in terms of power consumption, form factor and cost \cite{Hornibrook2015,conway2016fpga,Homulle2016, Charbon2016, charbon2017cryocmos, patra2017cryocmos, Sebastiano2017,ryan2017hardware}. However, defining the specifications to design such electronics requires a comprehensive analysis of the impact of the electronics performance on the quantum computer.


Furthermore, solid-state qubits need to be cooled to deep cryogenic temperatures. When operating the control electronics at cryogenic temperatures to relax the wiring requirements between the cryogenic quantum processor and its controller, as proposed in \cite{Hornibrook2015, conway2016fpga,Homulle2016, Charbon2016, charbon2017cryocmos, patra2017cryocmos, Sebastiano2017}, the need for accurate specifications is even more severe. The power dissipation of such a cryogenic controller is limited by the cooling power of the cryogenic refrigerator. For existing fridges, this is only $\sim$1\,W at 4\,K and $<$1\,mW below 100\,mK \cite{batey2014new}. Although this could improve in the future, e.g.~by adopting custom-made refrigerators \cite{Tavian851586}, the power consumption of the controller is also expected to increase to serve an increasing number of qubits. To meet these cooling constraints, the power dissipated by the electronics must be minimized by optimally allocating the available power across the various components of the classical controller. However, carrying out such optimization also demands a clear understanding of the impact of each components on the quantum-computer performance.

Analyzing the impact of the controller on the quantum computer's performance  has been attempted previously, but only for specific aspects of the control signals, such as the effect of microwave phase noise \cite{Ball2016,raftery2017direct}. 
To bridge this gap, the work presented here aims to provide a comprehensive analysis of the effect of non-ideal circuit blocks in the classical controller on the qubit fidelity for all possible operations, i.e.~single-qubit gates, two-qubit gates and read-out \footnote{Initialization is assumed to be performed by relaxation or by read-out}. This includes the effect of signal inaccuracies in frequency, voltage and time domain, and covers static, dynamic, systematic and random errors. Besides providing a general method for deriving the electronics specifications, a case study  targeting a 99.9\,\% average gate fidelity is analyzed.
This is particularly relevant since the minimum error rate to reach fault-tolerant quantum computing is around 99\% for a complete error correction cycle, thus requiring a single-operation fidelity above 99.9\% for a typical cycle of 10 operations\cite{Fowler2012}.
Reaching a 99.9\% fidelity for all operations is currently an ambitious goal for all qubit platforms. However, our model can be directly applied to analyse specifications for any given fidelity. The specifications resulting from the case study are mapped onto existing room-temperature integrated circuits (IC) to assess the feasibility of a practical controller.

Although the proposed approach can be easily extended to any quantum technology, such as 
NMR \cite{vandersypen2000experimental,vandersypen2001experimental,vandersypen2005nmr}, ion traps \cite{cirac1995quantum,monz201114}, superconducting qubits \cite{nakamura1999coherent,makhlin2001quantum,kelly2015state} or nitrogen vacancies in diamond \cite{childress2006coherent}, we focus on the specific case of single-electron spin qubits. This qubit technology offers promising prospects for large-scale quantum computing, thanks to the long coherence times \cite{Schreiber2014,yoneda201799}, the fully electrical control \cite{Kawakami2014,Veldhorst2014}, and potential integration of the quantum processor with a classical controller on a single chip fabricated using standard microelectronic technologies \cite{vandersypen2016interfacing}. The results obtained for the single-qubit gates can be generalized to any qubit system where single-qubit rotations are performed by applying a signal with a frequency matching the energy level spacing between the $|0\rangle$ and $|1\rangle$ states, e.g.~for NMR\cite{vandersypen2000experimental,vandersypen2001experimental,vandersypen2005nmr}, ion traps\cite{cirac1995quantum,monz201114}, NV-centers in diamond\cite{childress2006coherent} and transmons \cite{nakamura1999coherent,makhlin2001quantum,kelly2015state}. Similarly, the results obtained for the two-qubit gates can be generalized to any qubit system that exploits the exchange gate.

The paper is organized as follows: Section \ref{sec:system} describes the generalized spin-qubit quantum computer analyzed in this paper; Section \ref{sec:method} introduces the method for deriving the fidelity for the various operations; In Sections \ref{sec:1qubit}, \ref{sec:2qubit} and \ref{sec:readout} the electrical specifications required for single-qubit operations, two-qubit operations, and qubit read-out are derived, respectively. A discussion regarding the feasibility of these specifications follows in Section \ref{sec:discussion}. Conclusions are drawn in Section \ref{sec:conclusion}.

\section{\label{sec:system}A System-Level View of a Quantum Computer}

\subsection{The quantum processor}
\begin{figure}
\includegraphics[trim={0.7cm 1cm 0.7cm 1cm},clip,width=1.0\linewidth]{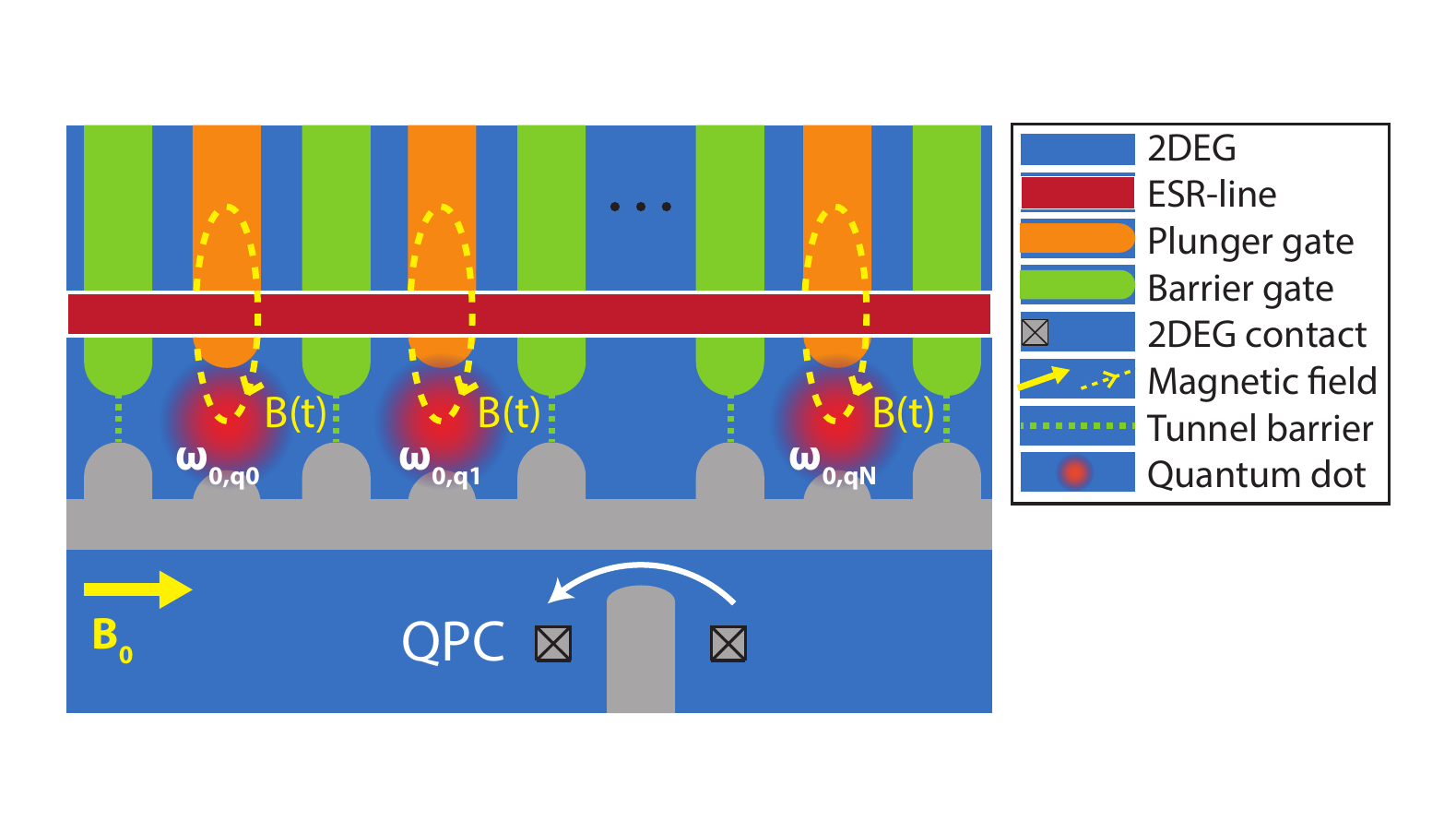}
\caption{\label{fig:qplane} A generic model of a spin-qubit quantum processor comprising qubits encoded in the spin of electrons trapped in quantum dots and a charge sensor (e.g. a QPC). The blue background indicates the 2DEG where locally quantum dots are formed, shown in red. Individual control over the dot potential and the tunnel barriers are assumed, using plunger gates (orange) and barrier gates (green), respectively. Furthermore, each qubit can have a unique Larmor frequency ($\omega_{0,qi}$).}
\end{figure}

A single-electron spin qubit encodes the quantum state in the spin state of a single electron. A generic model of a quantum processor based on single-electron spin qubits is shown in Fig.\,\ref{fig:qplane}, which captures all properties relevant for the interaction with the controller. Moreover, the figure illustrates a linear array of quantum dots, but this can be extended to more complex geometries like a 2D grid of quantum dots as shown in \cite{veldhorst2016silicon, vandersypen2016interfacing, li2017crossbar}.

Quantum dots are formed using a set of gate electrodes that locally deplete a 2-dimensional electron gas (2DEG) on a semiconductor chip (e.g.~a GaAs/AlGaAs heterojunction, a Si/SiGe heterojunction or a Si-MOS structure \cite{hanson2007spins, zwanenburg2013silicon}). Due to the small size of the quantum dot, the charge states become discrete with an energy level spacing related to the dot charging energy, thereby setting the required increase of the dot potential to add an electron to the dot. The dot potential, and thereby the number of electrons in the dot, is controlled by the plunger gate that capacitively couples to the quantum dot. Without loss of generality for the analysis of the electrical control signals, the availability of additional tunnel barrier gates is assumed which form tunnel barriers between neighboring dots by controlling the width of the depletion layer, thus allowing tunneling of electrons from and to the quantum dot in a tunable way. Early integration schemes involved non-overlapping gates (as shown in Fig.\,\ref{fig:qplane}) \cite{petta2005coherent,Kawakami2014,Kawakami2016}, while, in order to create better tunability and control, architectures now often include overlapping gates \cite{angus2007gate,Zajac2016,Veldhorst2014,Veldhorst2015}. 
The effect of crosstalk between different gates or the ESR-line is considered negligible or compensated for in the classical controller, and is not further discussed here, since it can be treated as a purely classical electrical effect.

An external static magnetic field $\mathbf{B_0}$ induces an energy difference between electrons with spin up and spin down, with Zeeman energy $E_z$. Because of the static magnetic field, the electron rotates around the Z-axis in the Bloch sphere with the Larmor frequency $\omega_0 = \gamma_e \cdot |\mathbf{B_0}|$, where $\gamma_e$ is the gyromagnetic ratio of the electron ($\gamma_{e} \approx$ 28 GHz/T in silicon). As indicated in Fig.\,\ref{fig:qplane}, each qubit can have a different Larmor frequency, which can be useful for the single-qubit or two-qubit operation \cite{Veldhorst2014,Veldhorst2015,Watson2017}\footnote{The Larmor frequency can be controlled by a magnetic field gradient, e.g.~using an on-chip micro-magnet \cite{tokura2006coherent, Kawakami2014, Kawakami2016,Watson2017} (not shown in Fig.\,\ref{fig:qplane}), or, alternatively, using the Stark shift, i.e.~by an electrical field provided with an extra confinement gate \cite{Veldhorst2014, Veldhorst2015} (not shown in Fig.\,\ref{fig:qplane}).}.

Single-qubit operations (Section \ref{sec:1qubit}) require the application of a varying magnetic field perpendicular to $\mathbf{B_0}$ and oscillating at the Larmor frequency.
In case of electron spin resonance (ESR), such a field is generated by a varying current in a nearby ESR-line \cite{Veldhorst2014, Veldhorst2015, Dehollain2013}. Alternatively, the same effect can be obtained e.g.~by applying a varying electric field to the electron in a spatial magnetic field gradient, as is the case for electric dipole spin resonance (EDSR)\cite{tokura2006coherent, Kawakami2014, Kawakami2016, Maurand2016}. The electric field variations are generated by a voltage on a nearby gate, for example through the plunger gate, without requiring an ESR-line.
Two-qubit operations (Section \ref{sec:2qubit}) and qubit read-out (Section \ref{sec:readout}) can be performed by pulsing the barrier and plunger gates.

Qubit read-out relies on a spin-to-charge conversion, followed by the detection of the eventual electron movement\cite{hanson2007spins}, using either a gate dispersive read-out \cite{colless2013dispersive}, or an additional charge sensor. The latter is assumed in this paper. For such charge sensor, different sensing techniques can be used, e.g.~a quantum point contact (QPC) \cite{reilly2007fast, vink2007cryogenic, elzerman2004single} or a single electron transistor (SET) \cite{Barthel2010}. As an example, Fig.\,\ref{fig:qplane} shows a QPC in close proximity to the quantum dots.

\subsection{The classical electronic controller}

\begin{figure}
\includegraphics[width=\linewidth]{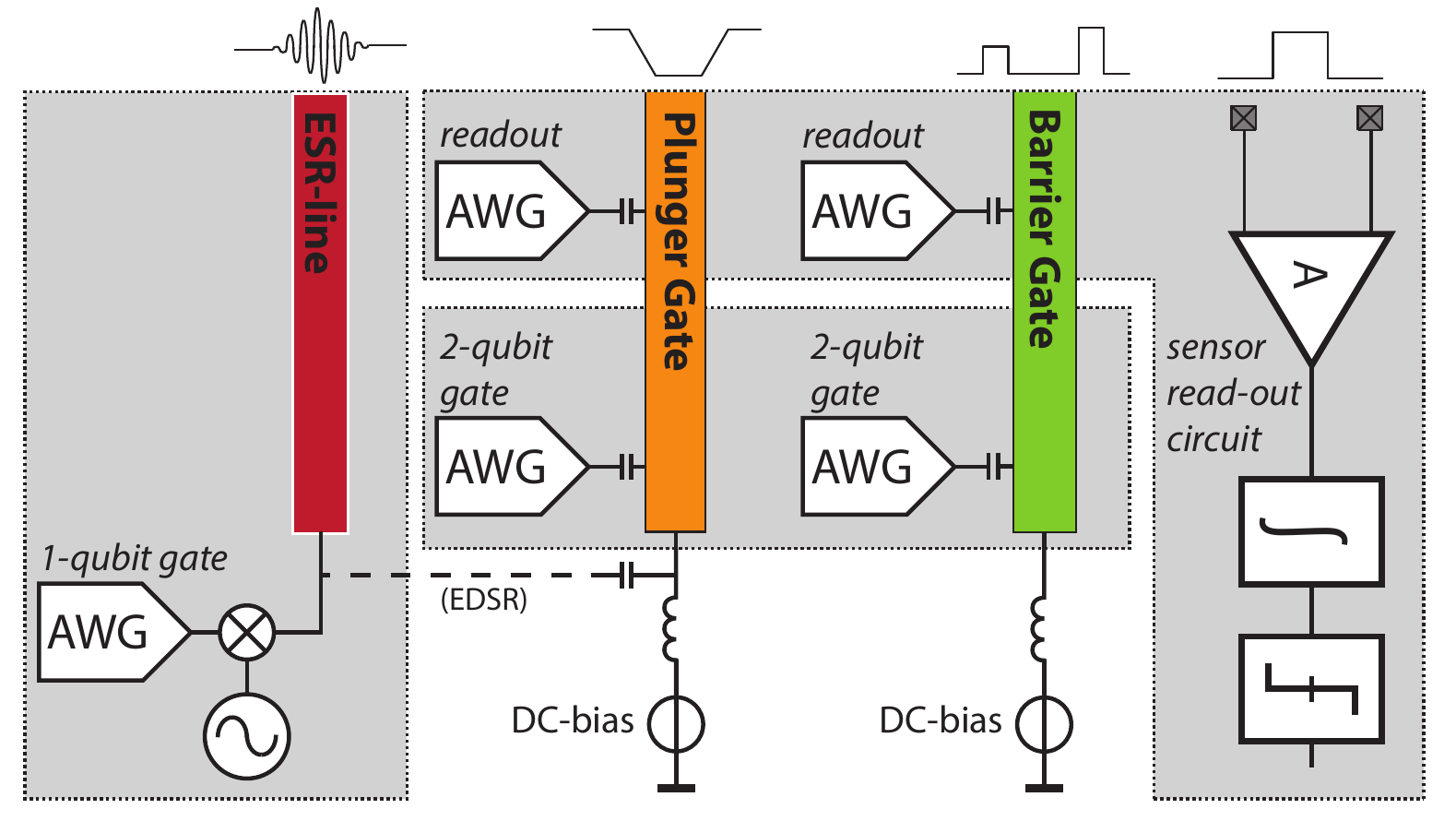}
\caption{\label{fig:electronics} The classical control electronics required for each line type (ESR-line, plunger gate and barrier gate) of the quantum processor shown in Fig.\,\ref{fig:qplane}. The electronics are shown as they are required for each of the possible operations, i.e.~single-qubit operation, two-qubit operation and read-out.}
\end{figure}

The classical controller is responsible for generating the required electrical signals to bias and control each gate and the ESR line, and for reading the state of the charge sensor. The required electronics have been schematically summarized in Fig.\,\ref{fig:electronics}.

When no operation is performed, each quantum dot must contain a single electron at the same dot potential and the tunnel barriers must be tuned to ensure a negligible coupling between neighbouring dots (Section \ref{sec:2q_nop}). Such conditions are ensured by the use of bias voltage generators as shown in Fig.\,\ref{fig:electronics}.

The oscillating magnetic field $B(t)$ required for single-qubit operations can be generated by an oscillating current $I(t)$, following the relation $B(t) = \alpha_I \cdot I(t)$ (ESR), or by an oscillating voltage $V(t)$, resulting in $B(t) = \alpha_V \cdot V(t)$ (EDSR).
The conversion factors $\alpha_I$ and $\alpha_V$ are influenced by many factors, such as the exact geometry of the structures, and can be determined experimentally.


This microwave current or voltage is generated by modulating a carrier from a local oscillator (LO) with an envelope produced by an arbitrary waveform generator (AWG). Although different hardware implementations are possible, this allows, without loss of generality, to split the carrier specifications, i.e.~the local oscillator specifications, from the envelope specifications, i.e.~the AWG specifications. In case each qubit has a unique Larmor frequency, a single control line could be used to control multiple qubits independently via frequency-division multiple access (FDMA), thus simplifying the wiring (Section \ref{sec:1q_fdma}). 

The voltage pulses required for the two-qubit gates and read-out are generated by AWGs. Distinct AWGs are assumed for two-qubit gates and read-out since the specifications for such operations can be different.

Besides the presented control electronics, additional hardware is required to process the signal from the charge sensor. The required hardware depends on the read-out method employed, e.g.~a direct measurement \cite{vink2007cryogenic} or RF-reflectometry \cite{reilly2007fast,colless2013dispersive}. As an example, a direct read-out, requiring a read-out amplifier, is shown in Fig.\,\ref{fig:electronics}.

\section{\label{sec:method} The Method of Analyzing the Effects of Signal Non-Idealities }
The evolution of the qubit state is evaluated by computing the system Hamiltonian ($H$), which is a function of the electrical signals applied by the classical controller. For static control signals, the Hamiltonian is time-independent and the unitary operation describing the evolution after a time $T$ is trivially $U = e^{-i \cdot H \cdot T}$ ($\hbar = 1$).

For dynamic signals, such as for complex signal envelopes
, the operation described by the time-varying Hamiltonian $H(t)$ is approximated by the product of time-independent components, leading to:
\begin{equation}
U \approx \prod_{n=N}^0  e^{-i \cdot H(n \cdot \Delta t) \cdot \Delta t},
\end{equation}
where $\Delta t$ is the time step that must be chosen small enough for the required accuracy of the approximation.

As a benchmark to evaluate how close $U$ is to the operation from an ideal controller $U_{ideal}$, the process fidelity is computed as \cite{nielsen2002simple, PEDERSEN2007}:
\begin{equation}
\label{eq:meth_static}
F = \frac{1}{n^2}\cdot\left|\text{Tr}\left[U_{ideal}^\dag U\right]\right|^2,
\end{equation}
for the $n$-dimensional complex Hilbert space ($n = 2$ for the single-qubit gate and $n = 4$ for the two-qubit gate).

In case of random dynamic errors, the ensemble average over all realizations has been evaluated, following \cite{green2012high,Green2013}. For a static but random error $\Delta$ for which $F = 1-c\cdot\Delta^2$, the expected fidelity simply follows as $F = 1-c\cdot\sigma^2$, if $\Delta$ follows a Gaussian distribution with standard deviation $\sigma$ and zero mean (see Appendix \ref{A:method}).

When treating random noise, the noise spectrum is relevant as the operation can be affected differently by noise at different frequencies. The method presented in \cite{green2012high,Green2013} is used to evaluate the expected process fidelity, and is outlined in Appendix \ref{A:1qubit_filt}.


\section{\label{sec:1qubit} Signal Specifications for Single-Qubit Operations}
As explained in Section \ref{sec:system}, the qubit rotates around the Z-axis due to the applied external magnetic field. Using an LO tuned at a frequency equal to the qubit's Larmor frequency, the qubit phase can be tracked and the qubit appears stationary in the reference frame of the LO. In this rotating frame, Z-rotations by a rotation angle $\theta_Z$ can easily be obtained by halting the LO for a duration $T = \theta_Z / \omega_{0}$, thereby acquiring the required qubit phase with respect to the LO. Equivalently, instead of halting the LO, the LO's phase can be instantaneously updated in software by an angle $\theta_Z$ \cite{vandersypen2001experimental,vandersypen2005nmr}.

For such a software defined Z-rotation, only the accuracy of the phase update of the LO matters. This is limited by the finite resolution in the phase setting. A phase error $\Delta\phi = \Delta\theta_Z$ reduces the fidelity of the Z-rotation as:
\begin{equation}
\label{eq:1_acc_z}
F_Z = 1 - \frac{1}{4} \cdot \Delta\phi^2.
\end{equation}

In the remainder of this section, we will focus on rotations around the X/Y-axis that 
are obtained by applying a magnetic field $B(t)$ oscillating at the qubit Larmor frequency $\omega_0$ and with specific phase, which is generated by applying either a microwave current or a microwave voltage as explained in Section \ref{sec:system}.

The Hamiltonian describing a single electron under microwave excitation in the lab frame is given by ($\hbar = 1$):
\begin{equation}
\label{eq:1_ham}
H_{lab} = -\omega_0 \cdot \frac{\sigma_z}{2} + \gamma_e \cdot B(t) \cdot \frac{\sigma_x}{2},
\end{equation}
where, here and in the following, $\sigma_x$, $\sigma_y$ and $\sigma_z$ are the Pauli matrices. The microwave magnetic field can be described as $B(t) = 2/\gamma_e \cdot \omega_R(t) \cdot \cos(\omega_{mw} t + \phi)$.

In the rest of this section, when deriving the various specifications, a constant amplitude ($\omega_R(t) = \omega_R$), i.e.~a rectangular envelope, is considered, unless stated otherwise. In the case of a rectangular envelope, $\omega_R$ is the Rabi frequency, i.e.~the rotation speed for the single-qubit gate. Note that for more complex envelopes, the resulting specifications for the control electronics can differ as the sensitivity to certain control parameters can be reduced when employing quantum optimum control, like GRAPE \cite{khaneja2005optimal}.

In the rotating wave approximation (RWA), the Hamiltonian in the frame rotating with  frequency $\omega_{mw}$ is:
\begin{equation}
\label{eq:1_ham_rot}
H \approx (\omega_{mw} - \omega_0) \cdot \frac{\sigma_z}{2} + \omega_{R} \cdot \left[ \cos(\phi) \cdot \frac{\sigma_x}{2} - \sin(\phi) \cdot \frac{\sigma_y}{2} \right].
\end{equation}

\begin{figure}
\includegraphics[width=\linewidth]{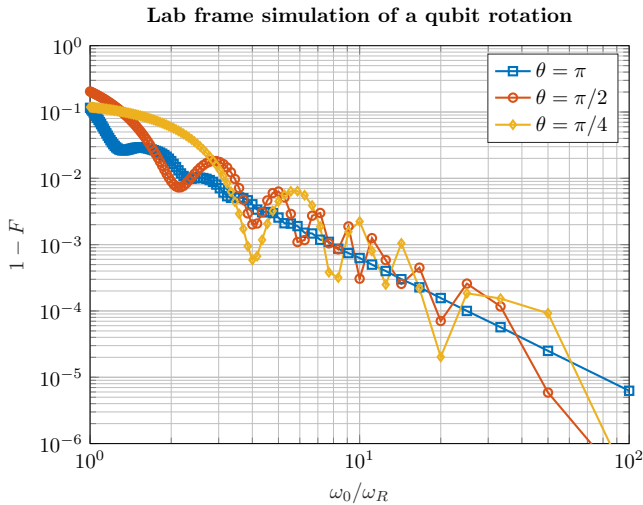}
\caption{\label{fig:1_rwa} The numerically simulated fidelity of a qubit rotation around the X-axis as a function of the $\omega_0/\omega_R$-ratio, without using the rotating wave approximation, for various rotation angles $\theta$.}
\end{figure}

This approximation is used to derive most specifications. However, the RWA is only valid when the Rabi frequency is sufficiently lower than the Larmor frequency ($\omega_R \ll \omega_0$). Fig.\,\ref{fig:1_rwa} shows the fidelity of a qubit rotation as obtained from a numerical simulation of the full Hamiltonian (Eq.\,\ref{eq:1_ham}). The fidelity has been computed with respect to the ideal rotation that would result from the Hamiltonian in the RWA (Eq.\,\ref{eq:1_ham_rot}). This plot clearly shows that a sufficiently high ratio of Larmor frequency to Rabi frequency is required for this approximation to hold. 

From Eq.\,\ref{eq:1_ham_rot} it follows that the rotation axis is affected by the matching of the microwave frequency ($\omega_{mw}$) to the Larmor frequency ($\omega_{0}$) and by the phase of the microwave signal ($\phi$), i.e.~the carrier signal. The rotation angle ($\theta = \omega_{R} \cdot T$), on the other hand, is determined by the amplitude of the signal ($\omega_{R}$) and the duration for which the microwave signal is applied ($T$), i.e.~the signal envelope. The effect of errors, both static and dynamic, on each of these parameters is analyzed in the subsequent sections. Detailed derivations of the formulas can be found in Appendix \ref{A:1qubit}.

\subsection{Specifications for the microwave carrier}
Although the microwave frequency should ideally be matched to the Larmor frequency, the carrier generating oscillator (Fig.\,\ref{fig:electronics}) will show a frequency inaccuracy $\Delta\omega_{mw} = \omega_{mw} - \omega_{0}$ due to finite frequency resolution and drift, which will lead to an undesired Z-component in the rotation 
according to Eq.\,\ref{eq:1_ham_rot}. The resulting fidelity is:
\begin{equation}
\label{eq:1_acc_f}
F_{X,Y} = 1 - \frac{1}{2} \cdot \left(\frac{\Delta\omega_{mw}}{\omega_R}\right)^2 \cdot \left[ 1 - \cos( \theta ) \right],
\end{equation}
where $\theta$ is the intended rotation angle, ranging from $-\pi$ to $\pi$.

For a given $\theta$, a better fidelity is achieved for a larger Rabi frequency, i.e.~a larger microwave amplitude and a shorter pulse duration. For instance, to achieve a $\pi$-rotation with 99.9\,\% fidelity, the required relative frequency accuracy $\Delta\omega_{mw}/\omega_R \approx 3\,\%$, translating to a frequency accuracy of 3\,kHz for a 100\,kHz Rabi frequency and 0.3\,MHz for a 10\,MHz Rabi frequency.

The axis of rotation is not only influenced by the microwave frequency, but also by the microwave phase (Eq.\,\ref{eq:1_ham_rot}).
Inaccuracies in the phase limit the fidelity according to:
\begin{equation}
\label{eq:1_acc_phi}
F_{X,Y} = 1 - \frac{1}{2} \cdot \Delta\phi^2 \cdot \left[ 1 - \cos( \theta ) \right].
\end{equation}
For instance, for a $\pi$-rotation with 99.9\,\% fidelity, the phase difference should be accurate to within $\Delta\phi \approx$ 0.03~rad or, equivalently, 1.8~$^{\circ}$. Reaching a certain phase accuracy $\Delta\phi$ is harder at higher LO frequencies, which might be required when operating at higher $\omega_R$ to fulfill the RWA requirement.
\\
\\
The matching of the microwave frequency and the qubit Larmor frequency is further limited by dynamic changes in either frequency. On the one hand, the Larmor frequency will drift due to e.g.~the hyperfine interaction with nuclear spins \cite{assali2011hyperfine}. In case such drift occurs on a timescale larger than the operation time, 
it can be considered a random static error and, hence, Eq.\,\ref{eq:1_acc_f} can be applied. On the other hand, the microwave frequency will vary due to phase noise of the oscillator. Although the effect of phase noise has already been studied in \cite{Ball2016}, a more realistic oscillator noise model, including both phase noise and wideband additive noise, 
has been adopted in this work. Deriving the intrinsic qubit filtering function for each noise contribution leads to an improved estimation of the fidelity that deviates from \cite{Ball2016}, as elaborated in the following.

The fidelity due to frequency noise in the microwave carrier is (Appendix \ref{A:1qubit_filt}):
\begin{equation}
\label{eq:1_filt_mw}
F_{X,Y} = 1 - \frac{1}{\pi} \int_{\omega_{min}}^{\infty} \frac{S_{mw}(\omega)}{\omega_R^2}
\cdot \left| H_{mw}(\omega) \right|^2 \cdot d\omega,
\end{equation}
where $S_{mw}(\omega)$ is the frequency noise power spectral density (PSD) and $\left| H_{mw}(\omega) \right|^2$ is the intrinsic qubit filter function, meaning that the qubit has a different sensitivity to noise at different frequencies. The lowest frequency of interest ($\omega_{min}$) is inversely proportional to the execution time of the quantum algorithm \footnote{Note that some quantum algorithms, such as dynamical decoupling sequences or error correction codes, can act as a high-pass filter for the noise thereby setting $\omega_{min}$.}.

\begin{figure}
\includegraphics[width=\linewidth]{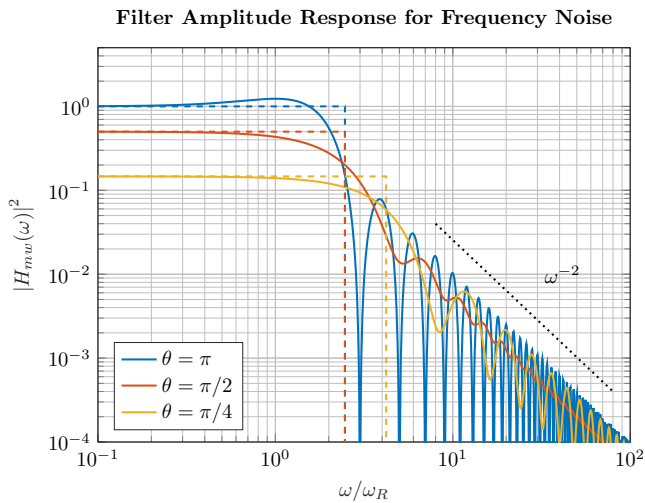}
\caption{\label{fig:1_filter_carrier_mw} The amplitude response of the intrinsic qubit filter for frequency noise, and its brick wall approximation (the dashed lines), for various rotation angles $\theta$.}
\end{figure}

The amplitude response of the filter function $\left| H_{mw}(\omega) \right|^2$ is shown in Fig.\,\ref{fig:1_filter_carrier_mw} for various rotation angles (for a full derivation refer to Appendix \ref{A:1qubit_filt}). This response represents a low-pass filter characteristic with a DC gain and effective noise bandwidth (ENBW) of:
\begin{eqnarray}
\label{eg:1_dc_mw}
\left| H_{mw}(0) \right|^2 &=& \frac{1}{2} \cdot \left[ 1 - \cos(\theta)\right]
\\
\label{eg:1_enbw_mw}
ENBW_{mw} &=& \omega_R \cdot \frac{\pi \left|\theta\right|}{2\left[ 1 - \cos(\theta)\right]},
\end{eqnarray}
which are indicated in Fig.\,\ref{fig:1_filter_carrier_mw} as the brick wall approximation of the filter. 

Note that the ENBW is proportional to the Rabi frequency, indicating that for faster operations, noise in a wider band affects the qubit. However, the lower limit of integration in Eq.\,\ref{eq:1_filt_mw} is also related to the operation time, as the lowest observable frequency is set by the duration of the quantum algorithm, which in turn scales with the time required for a qubit operation. In case of white noise, a good approximation is obtained with $\omega_{min} = 0$. Due to the factor $\frac{1}{\omega_R^2}$ in Eq.\,\ref{eq:1_filt_mw}, it is again advantageous to use the highest possible Rabi frequency. In case of flicker noise, the same conclusion holds, as then a higher $\omega_{min}$ is desirable.



\begin{figure}
\includegraphics[width=\linewidth]{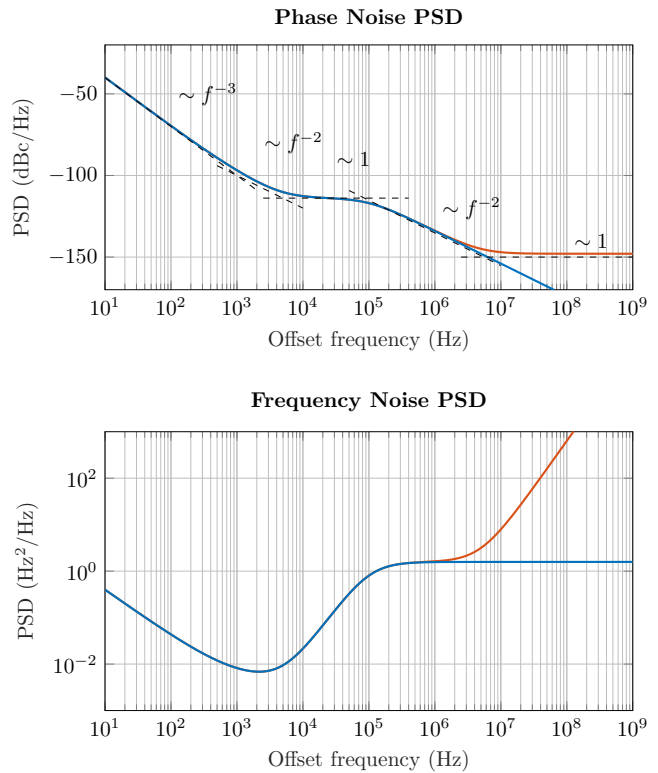}
\caption{\label{fig:1_psd} A typical plot of the phase noise and resulting frequency noise PSD of a PLL-based frequency generator. The red line indicates the noise as measured by a phase noise analyzer, whereas the blue line indicates the part of the noise that is actually phase noise. At high offset frequencies, where the lines diverge, wideband additive noise shows up in the phase noise plot, giving rise to a noise floor of around -150 dBc/Hz in this example.}
\end{figure}

As an example, the typically reported plot for the phase noise of a phase-locked loop (PLL) based frequency generator is shown in Fig.\,\ref{fig:1_psd} \cite{johns2008analog}. The microwave frequency noise is related to the oscillator's phase noise by:
\begin{equation}
\label{eq:1_phase_freq}
S_\omega(\Delta\omega) = \Delta\omega^2 \cdot S_\phi(\Delta\omega),
\end{equation}
where $S_\omega(\Delta\omega)$ and $S_\phi(\Delta\omega)$ are the frequency and phase noise PSD respectively at a frequency $\Delta\omega$ from the carrier $\omega_{mw}$, thus resulting in the frequency noise PSD as plotted in Fig.\,\ref{fig:1_psd}.
At low frequencies, the phase noise is typically limited by the flicker noise of the reference clock (the $\sim f^{-3}$ part). In the plot of the frequency noise PSD, this has a $f^{-1}$ roll-off, making it important to maximize $\omega_{min}$ in Eq.\,\ref{eq:1_filt_mw}. This could be resolved by using dynamical decoupling schemes which introduce an additional high-pass filtering \cite{uhrig2007keeping,biercuk2009optimized,green2012high,Green2013}.

The part of the phase-noise plot highlighted in red may be a source of concern \cite{Ball2016}, as it results in a frequency noise increasing as $f^2$ that exactly cancels the roll-off of the intrinsic qubit filter (Fig.\,\ref{fig:1_filter_carrier_mw}), thus resulting in a diverging integral for the fidelity (Eq.\,\ref{eq:1_filt_mw}) in case no additional band-pass filtering is applied. However, the noise highlighted in red visible in the phase noise plot originates from thermal noise added to the microwave signal by e.g., the output driver of the microwave signal generator \cite{Keysight2014}\cite{johns2008analog}.
The additive noise, with generally a wide bandwidth, is more accurately modeled in the applied microwave signal as:
\begin{equation}
B(t) = \frac{2 \omega_R}{\gamma_e} \cdot \cos(\omega_{mw} t + \phi + \phi_n(t)) + B_{add}(t),
\end{equation}
%
where $B_{add}(t)$ represents the additive noise with PSD $S_{add}(\omega)$. The actual phase noise $\phi_n(t)$, indicated by the blue line in Fig.\,\ref{fig:1_psd}, is clearly bandwidth limited by the qubit filter function.

\begin{figure}
\includegraphics[width=\linewidth]{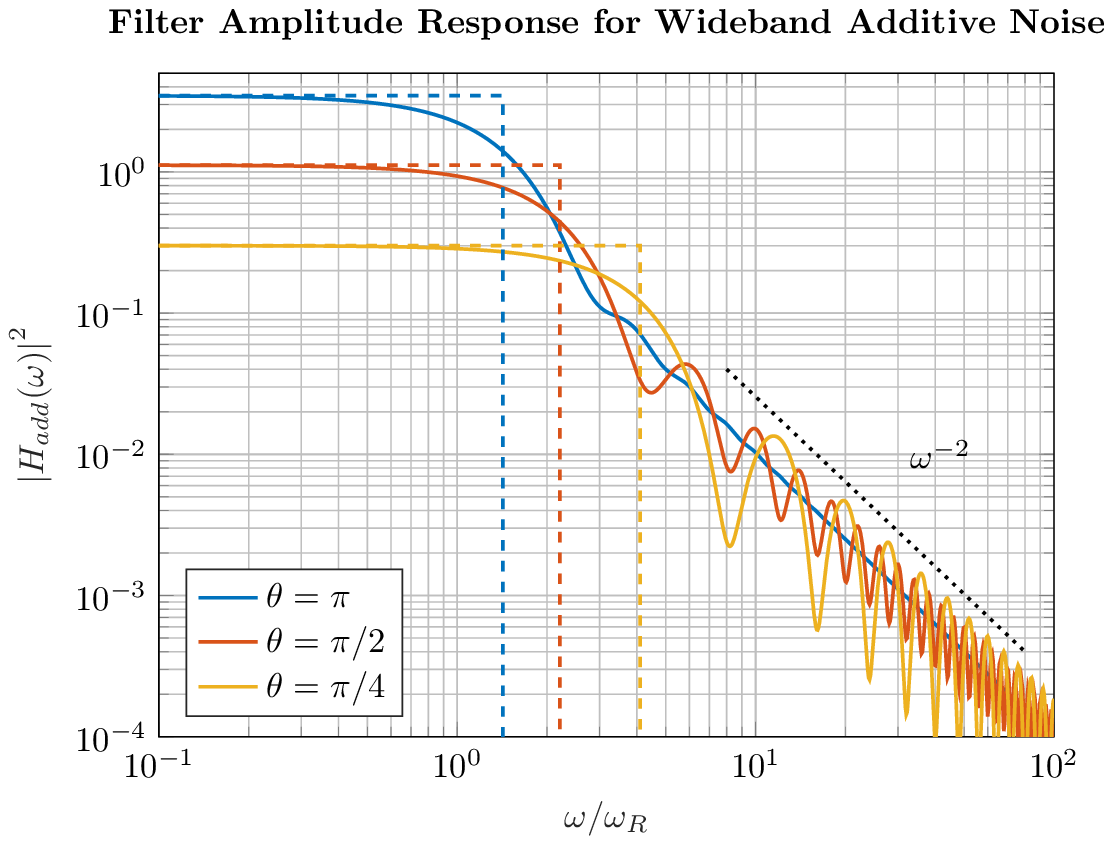}
\caption{\label{fig:1_filter_carrier_add} The amplitude response of the intrinsic qubit filter for wideband additive noise, and its brick wall approximation (the dashed lines), for various rotation angles $\theta$.}
\end{figure}

Unlike in \cite{Ball2016}, the effect of this additive noise on the qubit fidelity is evaluated separately. The PSD of this amplitude noise has the same frequency dependence as the PSD of the phase noise \cite{phillips2000noise}. The fidelity of the qubit operation in the presence of this type of noise can be derived as (Appendix \ref{A:1qubit_filt}):
\begin{equation}
F_{X,Y} = 1 - \frac{1}{\pi} \int_{\omega_{min}}^{\infty} \frac{S_{add}(\omega - \omega_0)}{\omega_R^2} \cdot \left| H_{add}(\omega) \right|^2  \cdot d\omega,
\end{equation}
where $\left| H_{add}(\omega) \right|^2$ can be interpreted as the amplitude response of the intrinsic qubit filter for this type of noise. A closed-form expression of $\left| H_{add}(\omega) \right|^2$ is given in Appendix \ref{A:1qubit_filt}. A plot of this amplitude response is shown in Fig.\,\ref{fig:1_filter_carrier_add} for various rotation angles, along with its brick wall approximation. As $\left| H_{add}(\omega) \right|^2$ has again a low-pass filter characteristic, the noise spectrum $S_{add}(\omega)$ is filtered by a band-pass filter centered around $\omega_0$. The DC gain and effective noise bandwidth (ENBW) of the low-pass filter characteristic are:
\begin{eqnarray}
\label{eg:1_dc_add}
\left| H_{add}(0) \right|^2 &=& \frac{1}{4} \cdot \theta^2 + \frac{1}{2} \cdot \left[ 1 - \cos(\theta)\right]
\\
\label{eg:1_enbw_add}
ENBW_{add} &=& \omega_R \cdot \frac{2\pi \cdot \theta}{\theta^2 + 2 \cdot [1 - \cos(\theta)]}.
\end{eqnarray}

\subsection{Specifications for the microwave envelope}
The microwave amplitude ($\propto \omega_R$) and duration ($T$) of the signal together determine the rotation angle ($\theta = \omega_R \cdot T$). Hence, any error in either one, again caused by the controller's finite resolution and drift, leads to an under- or over-rotation
, thereby reducing the fidelity:
\begin{eqnarray}
\label{eq:1_acc_theta}
F_{X,Y} &=& 1 - \frac{1}{4} \cdot \left( \frac{\Delta\omega_R}{\omega_R}\right)^2 \cdot \theta^2
\\
\label{eq:1_acc_duration}
F_{X,Y} &=& 1 - \frac{1}{4} \cdot \left( \frac{\Delta T}{T}\right)^2 \cdot \theta^2,
\end{eqnarray}
with $\Delta\omega_R/\omega_R$ and $\Delta T/T$ the relative error in the amplitude and duration, respectively.
In this case, targeting a $\pi$-rotation with 99.9\,\% fidelity, requires the relative error to be $ < 2\,\%$.

\begin{figure}
\includegraphics[width=\linewidth]{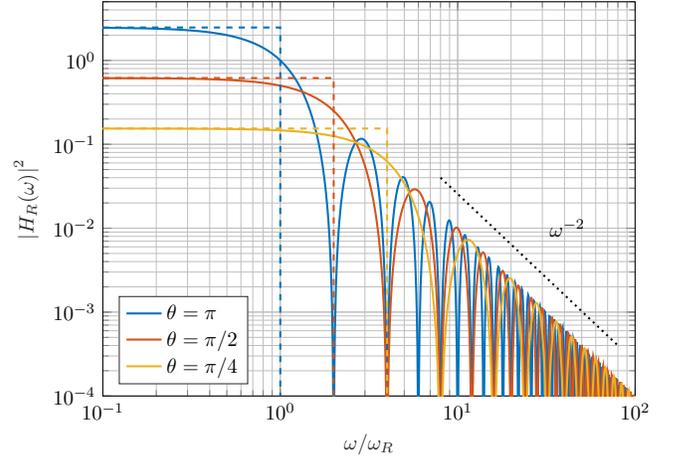}
\caption{\label{fig:qubit1_filter_envelope} The amplitude response of the intrinsic qubit filter for amplitude noise, and its brick wall approximation (the dashed lines), for various rotation angles $\theta$.}
\end{figure}

To evaluate the effect of dynamic variations in the amplitude with a noise PSD $S_{R}(\omega)$, the fidelity is evaluated as (Appendix \ref{A:1qubit_filt}):
\begin{equation}
\label{eq:1_filt_R}
F_{X,Y} = 1 - \frac{1}{\pi} \int_{\omega_{min}}^{\infty} \frac{S_{R}(\omega)}{\omega_R^2} \cdot \left| H_{R}(\omega) \right|^2 \cdot d\omega.
\end{equation}
Also for this type of noise, the noise PSD is filtered by an intrinsic qubit filter function with amplitude response $\left| H_{R}(\omega) \right|^2$, which again follows a low-pass filter characteristic with:
\begin{eqnarray}
\left| H_{R}(0) \right|^2 &=& \frac{1}{4} \cdot \theta^2
\\
ENBW_{R} &=& \omega_R \cdot \frac{\pi}{\left|\theta\right|}.
\end{eqnarray}
The closed-form expression of $\left| H_{R}(\omega) \right|^2$ is given in Appendix \ref{A:1qubit_filt} and a plot of the amplitude response with its brick wall approximation is shown in Fig.\,\ref{fig:qubit1_filter_envelope} for various rotation angles.

Just as for the frequency noise filtering, the frequency axis is defined relative to the Rabi frequency, i.e.~the bandwidth relevant for amplitude noise is proportional to the nominal amplitude. However, in case of white noise, Eq.\,\ref{eq:1_filt_R} simplifies to Eq.\,\ref{eq:1_acc_theta} with $\left( \Delta\omega_R/\omega_R\right)^2 = \sigma^2_{\omega_R}/\omega_R^2$ 
where $\sigma^2_{\omega_R}$ is the noise power. As can be seen, for a certain fidelity the required signal-to-noise ratio ($\omega_R^2/\sigma^2_{\omega_R}$) in the qubit's band of sensitivity is fixed.

Besides the microwave amplitude, also the signal duration $T$ is subject to random variations, i.e.~jitter. This period jitter is determined by the phase noise $S_{\phi}(f)$ of the reference clock used to set the duration, according to \cite{drakhlis2001calculate}\cite{johns2008analog}:
\begin{equation}
\label{eq:1_jitter}
\sigma_T = \frac{T}{\pi} \sqrt{\int_{f_{min}}^{\infty} S_{\phi}(f) \cdot \sin^2(2\pi f T) \cdot df}.
\end{equation}
This integral shows that the phase noise, which generally rolls-off with frequency (see Fig.\,\ref{fig:1_psd}), is filtered by a high-pass filter ($\sin^2(2\pi f T)$) with the corner frequency set by the duration $T$. The resulting variations in the duration, with standard deviation $\sigma_T$, lead to an infidelity that can be estimated using Eq.\,\ref{eq:1_acc_duration} with $\Delta T/T = \sigma_T/T$ assuming Gaussian distributed jitter.

\subsection{Requirements for qubit frequency multiplexing}
\label{sec:1q_fdma}

\begin{figure*}
    \subfloat[\label{fig_q1_fdma_env_1} The rectangular envelope under consideration.]{
        \includegraphics[width=.49\linewidth]{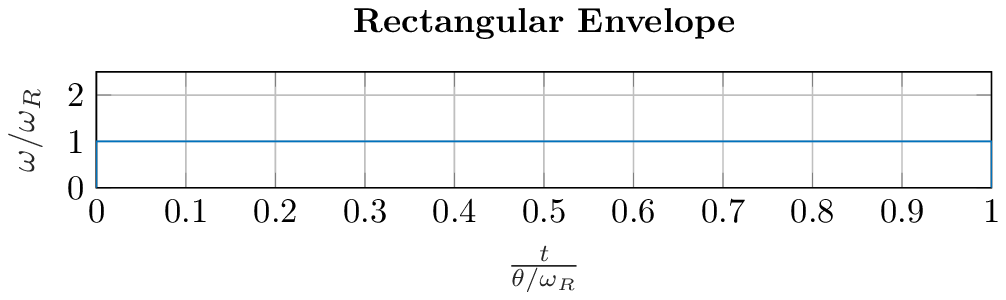}
    }
    \subfloat[\label{fig_q1_fdma_env_2} The Gaussian envelope under consideration.]{
        \includegraphics[width=.49\linewidth]{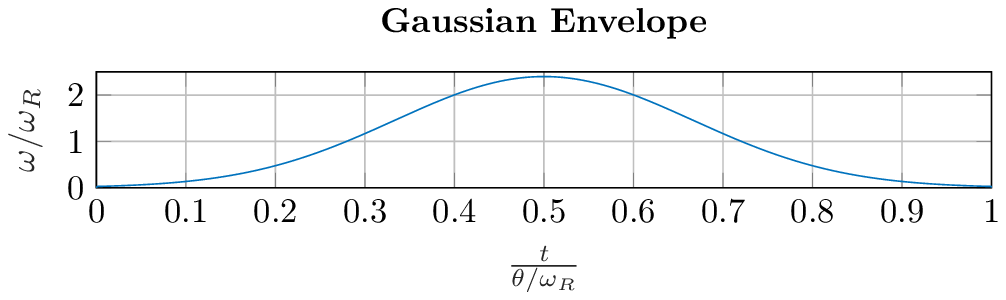}
    }
    \hfill
    \subfloat[\label{fig_q1_fdma_1} The infidelity of an identity operation (and amount of X/Y rotation and Z rotation) on a qubit spaced $\omega_{0,space}$ from the carrier for a rectangular envelope, along with the Fourier transform of the rectangular envelope.]{
        \includegraphics[width=.49\linewidth]{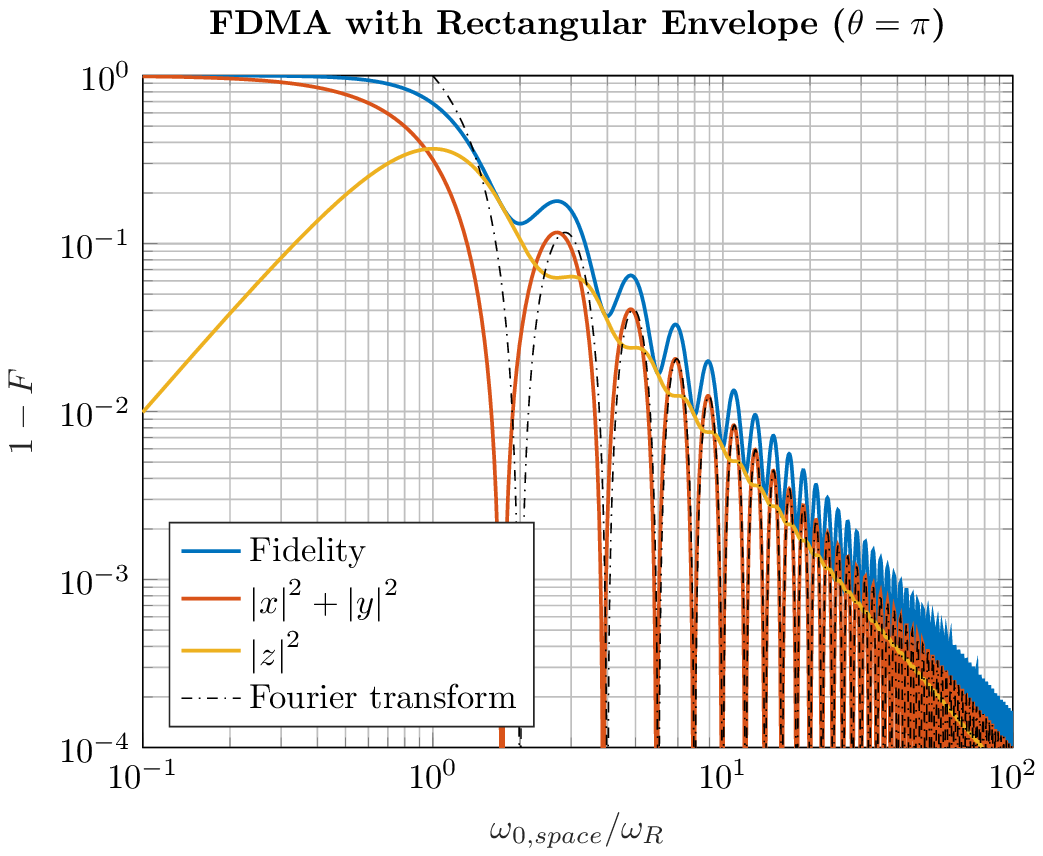}
    }
    \subfloat[\label{fig_q1_fdma_2} The infidelity of an identity operation (and amount of X/Y rotation and Z rotation) on a qubit spaced $\omega_{0,space}$ from the carrier for a Gaussian envelope, obtained by numerical simulation, along with the Fourier transform of the Gaussian envelope.]{
        \includegraphics[width=.49\linewidth]{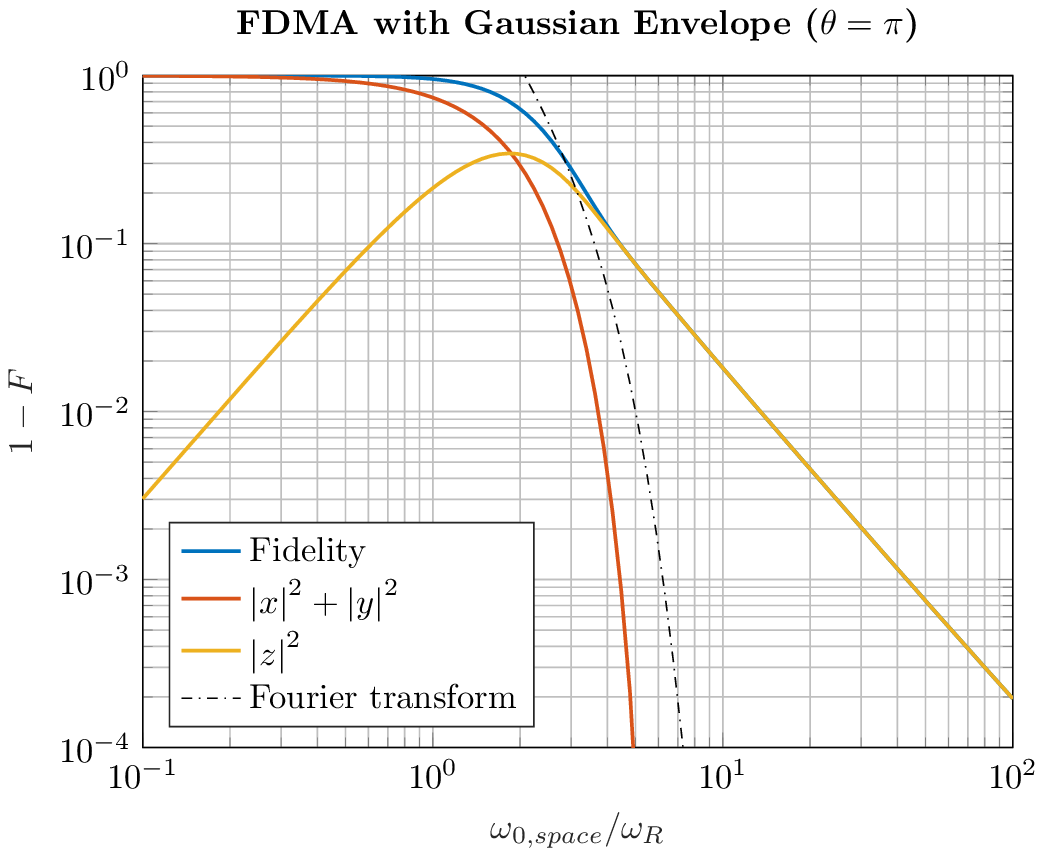}
    }
    \caption{\label{fig_q1_fdma_all}Qubit frequency multiplexing: envelopes and achievable fidelity.}
\end{figure*}

Multiple qubits sharing the same control line, i.e.~a single ESR line or control gates shorted together, can be controlled independently in case each qubit has a unique Larmor frequency, as mentioned in Section \ref{sec:system}. However, when rotating a qubit with Larmor frequency $\omega_0$ by applying a microwave signal at frequency $\omega_{mw} = \omega_0$, any unaddressed qubit on the same line with Larmor frequency $\omega_{0,other} = \omega_0 + \omega_{0,space}$ will be affected. Similarly, even if not on the same drive line, another qubit could be unintentionally driven due to parasitic coupling such as capacitive or magnetic crosstalk.

An expression for the fidelity of the unaddressed qubit with respect to the ideal identity operation is reported in Appendix \ref{A:1qubit_fdma} for a microwave pulse with rectangular envelope (Fig.\,\ref{fig_q1_fdma_env_1}) and it is plotted in Fig.\,\ref{fig_q1_fdma_1} where we assume the same Rabi frequency $\omega_R$ for both qubits. As expected, driving the qubit with a larger amplitude (i.e.~larger $\omega_R$) results in a shorter pulse for a given rotation angle, thus leading to a wider pulse bandwidth and, consequently, to a crosstalk extending to qubits further away in frequency.

Although the expectation may rise that reducing the pulse bandwidth by proper engineering of the pulse envelope can lead to a lower crosstalk, Fig.\,\ref{fig_q1_fdma_2} shows that also a Gaussian envelope (Fig.\,\ref{fig_q1_fdma_env_2}) does not result in a much faster roll-off. More insight in the reasons for this slow roll-off can be gained by decomposing the real operation into the identity operation $I$ and the Pauli operators $X$, $Y$ and $Z$ ($U_{real} = x\cdot X+y\cdot Y+z\cdot Z+i\cdot I$). Since $F_I=|i|^2 = 1-|x|^2 -|y|^2 - |z|^2 $ (Eq.\,\ref{eq:meth_static}, $U_{ideal} = I$), the fidelity is affected by contributions from unintended X, Y and Z-rotations. By individually looking at the rotations around the different axes, as shown in Figs.\,\ref{fig_q1_fdma_all}\subref{fig_q1_fdma_1},\subref{fig_q1_fdma_2}, it is clear that the fidelity is mainly limited by unintended Z-rotations of the unaddressed qubit for both the rectangular and Gaussian envelope. As shown in \cite{vandersypen2001experimental,vandersypen2005nmr}, these Z-rotations can be easily corrected by a shift of the microwave phase for subsequent operations on the unaddressed qubit. As this entails only a software update, the fidelity of those corrections is only limited by the phase accuracy of the microwave source as given by Eq.\,\ref{eq:1_acc_z}.

After applying the correction for the Z-rotation, the fidelity of the identity operation on the unaddressed qubit improves to (Appendix \ref{A:1qubit_fdma}):
\begin{equation}
\label{eq:1_fdma_fcorr}
F_{I_{(Z-corrected)}} \approx 1-\frac{\beta^2}{\alpha^2}\cdot\sin^2\left(\frac{\theta}{2}\alpha\right) \ge 1-\frac{\beta^2}{\alpha^2},
\end{equation}
where $\alpha = \frac{\omega_{0,space}}{\omega_R}$ and $\beta = \frac{\omega_{R,unaddressed}}{\omega_R}$, where in general the unaddressed qubit can have a different Rabi frequency ($\omega_{R,unaddressed}$) at the same microwave amplitude, e.g.~due to a lower coupling to the drive signal. As expected, the fidelity given by Eq.\,\ref{eq:1_fdma_fcorr} is proportional to the spectrum of the envelope of the applied pulse, both for the rectangular envelope (Fig.\,\ref{fig_q1_fdma_1}) and the Gaussian envelope (Fig.\,\ref{fig_q1_fdma_2}). Consequently, reducing the pulse bandwidth by proper engineering of the pulse envelope is an effective solution if the unintended Z-rotations are corrected.

\begin{figure*}
    \subfloat[\label{fig_q1_fdma_3} The frequency spacing required to achieve a certain fidelity at given relative signal strength $\beta$, for a rectangular envelope. The upper bound is given in Eq.\,\ref{eq:1_fdma_fcorr}.]{
        \includegraphics[width=0.49\linewidth]{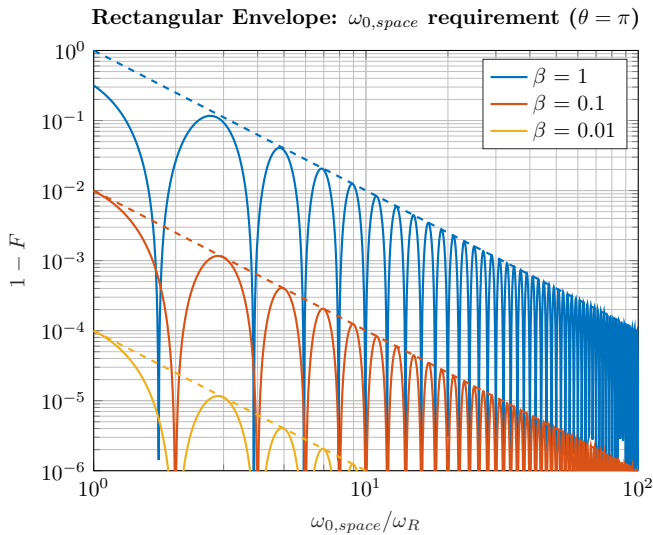}
    } 
    \subfloat[\label{fig_q1_fdma_4} The driving tone attenuation $\beta$ required at a certain frequency spacing to achieve a given fidelity, for a rectangular envelope. The lower bound is given in Eq.\,\ref{eq:1_fdma_fcorr_2}.]{
        \includegraphics[width=0.49\linewidth]{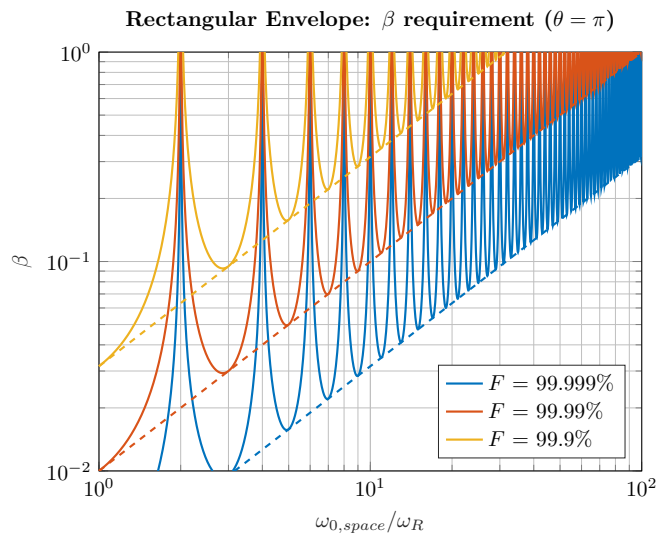}
    } 
    \caption{Qubit frequency multiplexing: requirements in case of a rectangular envelope.}
\end{figure*}

A certain minimum frequency separation is necessary to achieve a target fidelity, as shown in Fig.\,\ref{fig_q1_fdma_3} for the rectangular envelope. The lower bound on the fidelity as given in Eq.\,\ref{eq:1_fdma_fcorr} has been plotted as well, as the notches in the graph move depending on $\theta$. Similarly, if the coupling of the microwave drive is due to parasitic effects and it is unwanted, a target fidelity for unaddressed qubits translates into a requirement in the driving tone attenuation (Fig.\,\ref{fig_q1_fdma_4}):
\begin{equation}
\label{eq:1_fdma_fcorr_2}
\beta = \sqrt{1-F_{corr}} \cdot \frac{\alpha}{\left|\sin\left(\frac{\theta}{2}\alpha\right)\right|} \ge \sqrt{1-F_{corr}} \cdot \frac{\omega_{0,space}}{\omega_R}.
\end{equation}

Finally, FDMA has the potential to perform single-qubit gates on several qubits at the same time using a single drive line. In that case, it is not sufficient to apply a compensating Z-rotation afterwards on another qubit if that qubit is also performing an operation. As the Z-rotation is obtained gradually when an off-resonance tone is applied, the driving tone applied to perform the operation should be altered to compensate for this Z-rotation during the operation. This requires a proper engineering of all the microwave pulses that are applied simultaneously \cite{mccoy1993selective,KUPCE1995261,Pauly1991,steffen2000simultaneous,vandersypen2001experimental}.

\subsection{Specifications for the idle operation}
\label{sec:1q_nop}

\begin{figure}
\includegraphics[width=\linewidth]{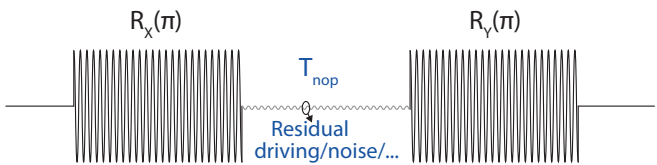}
\caption{\label{fig:idle_time} The state of a qubit is affected during idle times between operations due to e.g.~residual driving on the ESR-line.}
\end{figure}

In a typical quantum algorithm, a qubit can be idle for a while, waiting for the operations on other qubits to finish, before being operated on, e.g., due to limitations in the hardware or data dependencies. As discussed in Section \ref{sec:1q_fdma}, the state of a qubit is affected when driving another qubit on the same line. This section discusses other processes that cause the state of the qubit to degrade during an idle period lasting $T_{nop}$, as indicated in Fig.\,\ref{fig:idle_time}.

As evident from Eq.\,\ref{eq:1_ham_rot}, a qubit will perform a Z-rotation in the rotating frame if the microwave frequency is not matched to the qubit's Larmor frequency, even when the driving tone is not applied to the qubit. 
Evaluating the fidelity of an identity operation in case of a frequency inaccuracy $\Delta\omega_{mw,0}$ leads to:
\begin{equation}
\label{eq:1_off_omega}
F_I = 1 - \frac{1}{4} \cdot \Delta\omega_{mw,0}^2 \cdot T_{nop}^2.
\end{equation}
%
As an example, consider that the qubit is idle for 10 times the duration of a $\pi$-rotation of another qubit at a certain Rabi frequency. In order to obtain a 99.9\,\% fidelity, the frequency inaccuracy should be kept smaller than 0.2\,\% of that Rabi frequency. Note that this requirement on frequency accuracy due to the idle operation can easily be more stringent than the requirement due to a rotation (Eq.\,\ref{eq:1_acc_f}).

Besides Z-rotations, unintended X- and Y-rotations of the qubit are possible in case power is present at the qubit's Larmor frequency (Fig.\,\ref{fig:idle_time}). In general, a tone could be present at the qubit frequency, e.g.~due to signal leakage from the microwave source or non-linearities in the system leading to harmonic or inter-modulation tones. 
The presence of a spurious tone that would give a Rabi frequency of $\omega_{spur}$ will reduce the fidelity as:
\begin{equation}
F_I = 1 - \frac{1}{4} \cdot \omega_{spur}^2 \cdot T_{nop}^2.
\end{equation}
%
Considering the same example, in order to keep a 99.9\,\% fidelity, the residual driving tone needs to be 54\,dB lower than the amplitude used to perform a $\pi$-rotation.

Besides a tone, residual thermal noise could be present on the drive line. Consider the Hamiltonian in the lab frame (Eq.\,\ref{eq:1_ham}) with $\omega_{ESR}(t) = 2\cdot\omega_{R_n}(t)$, where $\omega_{R_n}(t)$ is the noise signal with spectral density $S_{R_n}(\omega)$. The fidelity is obtained as (Appendix \ref{A:1qubit_nop}):
%
\begin{equation}
%
F_I = 1 - \frac{1}{\pi} \int_{0}^{\infty} S_{R_n}(\omega) \cdot |H_n(\omega)|^2 \cdot d\omega,
\end{equation}
where
\begin{equation}
|H_n(\omega)|^2 = 2 \cdot \frac{\sin^2\left(\frac{T_{nop}}{2}\cdot(\omega-\omega_0)\right)}{(\omega-\omega_0)^2},
\end{equation}
which indicates that the noise spectrum is filtered by a sinc-shaped band-pass filter centered around $\omega_0$, with brick wall approximation:
\begin{equation}
|H_n(\omega)|^2 \approx
\begin{cases}
T_{nop}^2 / 2 & |\omega-\omega_0| \le \pi/T_{nop}\\
0 &\text{elsewhere}.
\end{cases}
\end{equation}
%

Besides the effects discussed in this section, the qubit state will be lost by interactions with other qubits, which is further discussed in Section \ref{sec:2q_nop}.

\subsection{Case study of the specifications for a single-qubit operation}
\label{sec:1q_specs}

\begin{table*}
  \centering
  \caption{Example specifications for the control electronics. The PSD values provided as a comment assume a white spectrum for the amplitude and frequency noise (i.e.~-20\,dB/dec for the phase noise).}
  \begin{ruledtabular}
    \begin{tabular}{lllll}
          & \multicolumn{1}{l}{\textbf{Value}} & \multicolumn{2}{c}{\textbf{Infidelity contribution}} & \multicolumn{1}{l}{\textbf{Comment}} \\
          &       & \textbf{to an operation} & \textbf{to idling} &  \\
    \hline
    \textit{\textbf{Frequency}} &       &       &       &  \\
    nominal & \multicolumn{1}{l}{10\,GHz} & $0.64\times 10^{-9}$ &       & \multicolumn{1}{l}{RWA when driving a qubit} \\
    spacing & \multicolumn{1}{l}{1\,GHz} &       & 1$\times 10^{-6}$ & \multicolumn{1}{l}{FDMA leakage with rectangular envelopes} \\
    inaccuracy & \multicolumn{1}{l}{11\,kHz} & 125$\times 10^{-6}$ & 308$\times 10^{-6}$ &  \\
    oscillator noise & \multicolumn{1}{l}{11\,kHz$_{rms}$} & 125$\times 10^{-6}$ & 308$\times 10^{-6}$ & \multicolumn{1}{l}{ENBW = 2.5\,MHz, $\mathcal{L}$(1\,MHz) = -106\,dBc/Hz} \\ 
    nuclear spin noise & \multicolumn{1}{l}{1.9\,kHz$_{rms}$} & 3.6$\times 10^{-6}$ & 8.9$\times 10^{-6}$ & \multicolumn{1}{l}{From \cite{Veldhorst2014}, T$_2^*$ = 120\,$\mu$s} \\
    wideband noise & \multicolumn{1}{l}{12\,$\mu$V$_{rms}$} & 125$\times 10^{-6}$ &  & \multicolumn{1}{l}{ENBW = 2.9\,MHz, $S_{add}$ = 7.1\,nV/$\sqrt{\text{Hz}}$} \\
    \hline
    \textit{\textbf{Phase}} &       &       &       &  \\
    inaccuracy & \multicolumn{1}{l}{0.64\,$^{\circ}$} & 125$\times 10^{-6}$ & 31$\times 10^{-6}$ & \multicolumn{1}{l}{FDMA Z-corrections limit the no operation} \\
    \hline
    \textit{\textbf{Amplitude}} &       &       &       &  \\
    nominal & \multicolumn{1}{l}{2\,mV} &       &       & \multicolumn{1}{l}{Full-scale: 4\,mV, RMS: 1.4\,mV$_{rms}$} \\
    inaccuracy & \multicolumn{1}{l}{14\,$\mu$V} & 125$\times 10^{-6}$ &       & \\
    noise & \multicolumn{1}{l}{14\,$\mu$V$_{rms}$} & 125$\times 10^{-6}$ &       & \multicolumn{1}{l}{ENBW = 1.0\,MHz, PSD = 14\,nV/$\sqrt{\text{Hz}}$, SNR = -40\,dB} \\
    off-spur   & \multicolumn{1}{l}{19\,$\mu$V} &       & 217$\times 10^{-6}$ & \multicolumn{1}{l}{-41\,dBc} \\
    off-noise   & \multicolumn{1}{l}{10\,$\mu$V$_{rms}$} &       & 125$\times 10^{-6}$ & \multicolumn{1}{l}{ENBW = 2.0\,MHz, PSD = 7.1\,nV/$\sqrt{\text{Hz}}$} \\
    \hline
    \textit{\textbf{Duration}} &       &       &       &  \\
    nominal & \multicolumn{1}{l}{500\,ns} &       &       &  \\
    inaccuracy & \multicolumn{1}{l}{3.6\,ns} & 125$\times 10^{-6}$ &       &  \\
    noise & \multicolumn{1}{l}{3.6\,ns$_{rms}$} & 125$\times 10^{-6}$ &       &  \\
    \hline
     &       & \textbf{$F_{X,Y}$ = 99.9\,\%} & \textbf{$F_{I}$ = 99.9\,\%} &  \\
    \end{tabular}%
  \end{ruledtabular}
  \label{tab:1_specs}%
\end{table*}%

With the information provided in this section, clear specifications for the control electronics can be derived. Table \ref{tab:1_specs} shows as an example how the total error budget can be allocated over the electronics specification to achieve a 99.9\,\% fidelity for a $\pi$-rotation at a Rabi frequency of 1\,MHz. The same fidelity is targeted for preserving the state of the qubit when not operating on it for a time equal to the operation time ($T_{nop} = T$). 
The example considers the use of simple rectangular pulses, without any echo technique.

A Larmor frequency larger than 80\,MHz would be sufficient not to get impaired by fast oscillating terms neglected by the RWA (see Fig.\,\ref{fig:1_rwa}). However, choosing $\omega_0$ = 10\,GHz is more in line with values used in practice, and allows for a large qubit frequency spacing. A frequency spacing of 1\,GHz was selected, the same as is considered in the case study of 2-qubit operations (Section \ref{sec:1q_specs}). Such spacing is however $\sim$10 times larger than required for FMDA crosstalk minimization.
%
The example also shows the effect of the frequency noise as expected from isotopically purified Si-28 (800\,ppm \textsuperscript{29}Si), highlighting that its contribution to the infidelity is negligible in this example.

The values provided for the microwave amplitude assume a qubit plane based on EDSR where an amplitude of 2\,mV at the gate is required for a Rabi frequency of 1\,MHz (close to the value reported in \cite{Kawakami2014}). All specifications are valid at the gate so that wiring attenuation and filtering might need to be factored in to refer the specifications back to the electronics.

Following these specifications, the microwave envelope (amplitude and duration) can be generated, for instance, by an AWG 
with a sample rate of at least 150\,MS/s, such that the sample time is less than 6.7\,ns, resulting in a maximum inaccuracy of 3.3\,ns. Furthermore, the AWG should have a resolution of 8\,bits, such that at a full-scale swing of 4\,mV, the quantization step is sufficiently low. To meet the noise requirement and the specifications on the residual driving when not operating the qubit (`off-spur' in Table \ref{tab:1_specs}), an effective number of bits (ENOB) of only 6.5\,bits is required.

The LO used for the up-conversion requires a frequency resolution of $\sim$ 20\,kHz (for the inaccuracy). Assuming a -20\,dB/dec slope of the phase noise, the single-side band phase noise at 1\,MHz from the carrier, $\mathcal{L}$(1\,MHz), needs to be below -106\,dBc/Hz. 
Furthermore, the LO's phase inaccuracy needs to be below 0.64\,$^{\circ}$.


\section{\label{sec:2qubit} Signal Specifications for Two-Qubit Operations}
As stated in Section \ref{sec:system}, by default the tunnel coupling between the qubits is negligible, and the qubits have the same potential, i.e.~they are not detuned. By increasing the tunnel coupling and/or by detuning the qubits, the qubit interaction increases and a two-qubit gate can be obtained. In this system, by leveraging this exchange interaction, a 2-qubit exchange gate and a C-phase gate can be implemented. With either of these gates, and single-qubit operations, a universal set is obtained.

To describe the physical interactions required for the two-qubit gate, higher energy levels need to be modeled in the Hamiltonian. The analysis presented here is limited to the interaction between two neighboring qubits and to the single-dot singlet states ($\left| 0,2 \right\rangle$ represents the singlet state in the right dot and $\left| 2,0 \right\rangle$ the singlet state in the left). In the basis $\Psi = \left[ \left| \uparrow, \uparrow \right\rangle, \left| \uparrow, \downarrow \right\rangle, \left| \downarrow, \uparrow \right\rangle, \left| \downarrow, \downarrow \right\rangle, \left| 0,2 \right\rangle, \left| 2,0 \right\rangle \right]$, the Hamiltonian of a double quantum dot is given by ($\hbar = 1$) \cite{coish2007exchange,meunier2011efficient,Veldhorst2015}:
\begin{equation}
\label{eq:2_ham}
H = \begin{bmatrix}
 -\omega_0  & 0 & 0 & 0 & 0 & 0 \\
 0 & \frac{\delta\omega_{0}}{2} & 0 & 0 & t_0 & t_0 \\
 0 & 0 & -\frac{\delta\omega_{0}}{2} & 0 & -t_0 & -t_0 \\
 0 & 0 & 0 & \omega_{0} & 0 & 0 \\
 0 & t_0 & -t_0 & 0 & U - \epsilon & 0 \\
 0 & t_0 & -t_0 & 0 & 0 & U + \epsilon \\
\end{bmatrix},
\end{equation}
%
where $\omega_{0} = (\omega_{0,A} + \omega_{0,B})/2$, $\delta\omega_{0} = \omega_{0,B} - \omega_{0,A}$ and $\omega_{0,A}$ and $\omega_{0,B}$ are the Larmor frequencies of the two qubits. 
The charging energy ($U$) is assumed to be the same for both dots. The tunnel coupling between the quantum dots ($t_0$) has an exponential relation to the voltage on the barrier gate, and the detuning energy ($\epsilon$) is controlled by the voltage difference on the plunger gates of the dots ($V_d$) via the lever arm $\alpha = \Delta\epsilon / \Delta V_d$. 

\begin{figure}
\includegraphics[width=\linewidth]{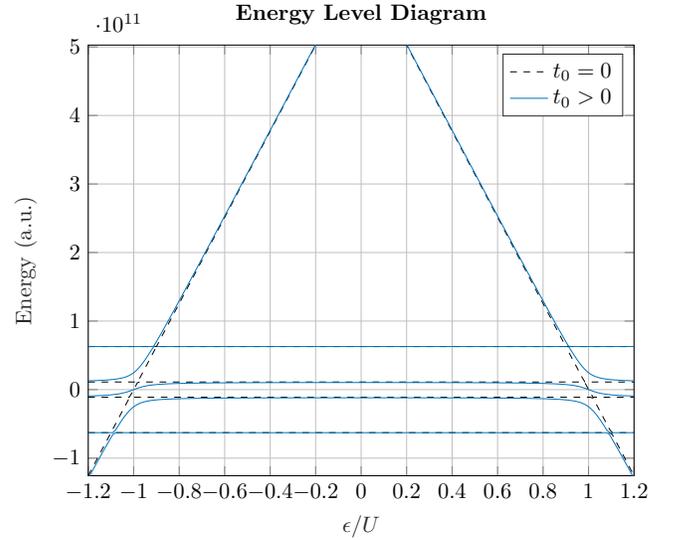}
\caption{\label{fig_q2_energydiagram} The energy level diagram of the 2-qubit system. An avoided crossing is visible for $|\epsilon| = U$ when there is a finite tunnel coupling between the dots.}
\end{figure}

An avoided crossing is observed in the energy level diagram for $|\epsilon| = U$ and $t_0 > 0$ (Fig.\,\ref{fig_q2_energydiagram}), that gives rise to eigenenergies different from the case of 2 isolated dots ($t_0 \sim 0$) for any detuning. This change of eigenenergy, and corresponding eigenstate, forms the basis of the two-qubit operations. An investigation of the eigenenergies ($\omega_{\lambda_i}$) of the Hamiltonian in Eq. \ref{eq:2_ham} reveals that the four relevant eigenenergies are given by (Appendix \ref{A:2qubit_lamba}, $\hbar = 1$):
\begin{eqnarray}
\omega_{\lambda_1} &=& -\omega_0\\
\omega_{\lambda_2} &\approx& \begin{cases}\frac{-\omega_{op} + \sqrt{\delta\omega_{0}^2 + \omega_{op}^2}}{2}  & 0 \le \delta\omega_{0} < \sqrt{2} t_0 \\ \frac{-\omega_{op} + \delta\omega_{0}}{2}  & \delta\omega_{0} = \sqrt{2} t_0 \end{cases}\\
\omega_{\lambda_3} &\approx& \begin{cases} \frac{-\omega_{op} - \sqrt{\delta\omega_{0}^2 + \omega_{op}^2}}{2} & 0 \le \delta\omega_{0} < \sqrt{2} t_0 \\ \frac{-\omega_{op} - \delta\omega_{0}}{2} & \delta\omega_{0} = \sqrt{2} t_0 \end{cases}\\
\omega_{\lambda_4} &=& \omega_0,
\end{eqnarray}
where
\begin{equation}
\label{eq:2_speed}
\omega_{op} = 4 \cdot t_0^2 \cdot \frac{U}{U^2 - \epsilon^2}.
\end{equation}

Note that the expression used in this paper for $\omega_{op}$ derives directly from the Hamiltonian of Eq.\,\ref{eq:2_ham}. Experiments have reported $\omega_{op}$ as an exponential function of detuning \cite{laird2006effect}.

\begin{figure}
\includegraphics[width=\linewidth]{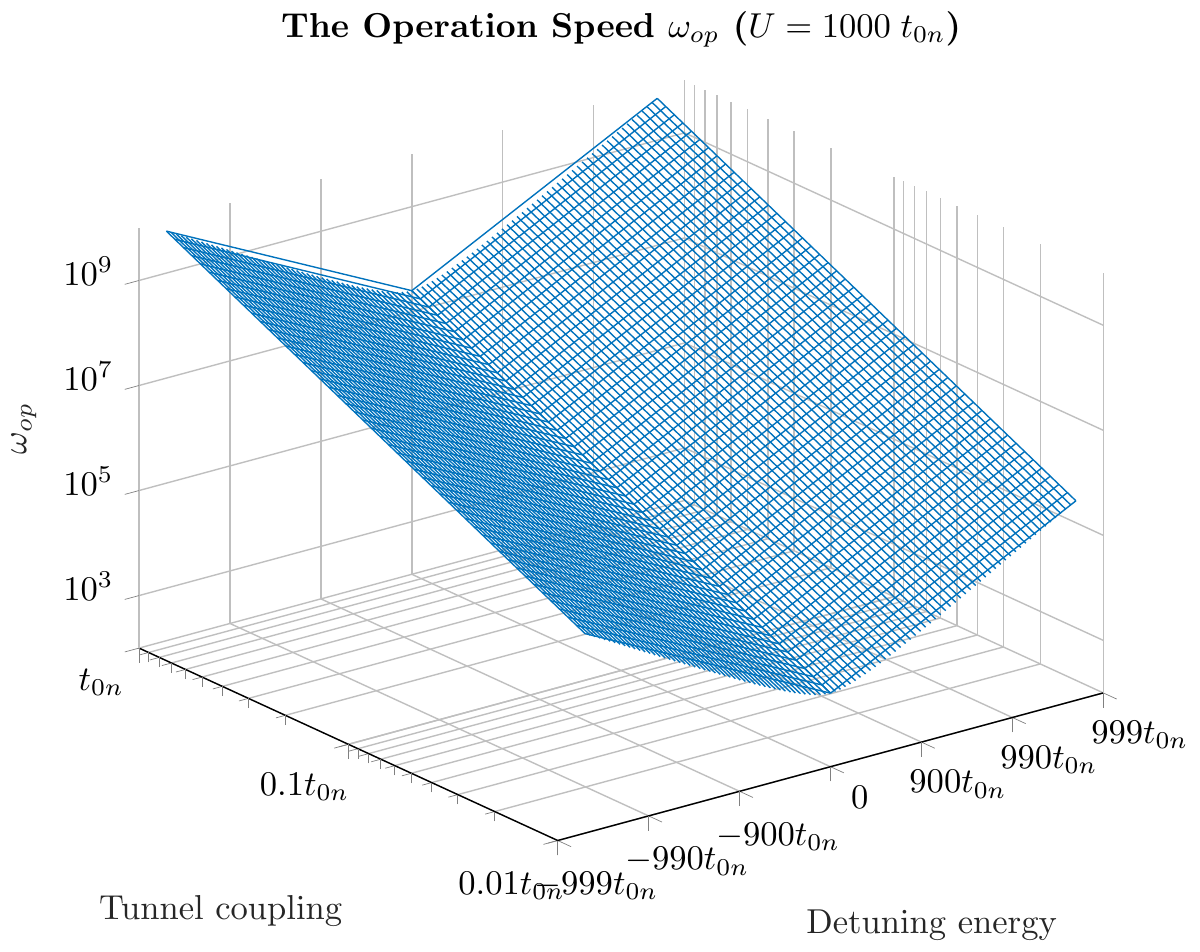}
\caption{\label{fig_q2_frequency} The 2-qubit operation speed $\omega_{op}$ (Eq. \ref{eq:2_speed}) versus the interdot tunnel coupling and detuning. A nominal tunnel coupling $t_{0n}$ of 1 GHz is used.}
\end{figure}

As $\omega_{op}$ describes the amount of exchange interaction, it directly sets the speed of the 2-qubit operation. A plot of $\omega_{op}$ versus the tunnel coupling and detuning is shown in Fig.\,\ref{fig_q2_frequency}. To perform the two-qubit operation, a control pulse must be applied, to move the system away from the default point (negligible tunnel coupling and zero detuning) to the desired operating point where there is sufficient exchange interaction such that a 2-qubit operation is performed. From Fig.\,\ref{fig_q2_frequency} it is clear that a fast gate can be obtained at finite detuning, becoming faster closer to the avoided crossing, controlled by the detuning and/or tunnel coupling. Alteratively, operation at zero detuning (the charge symmetry point \cite{reed2016reduced}) is possible, controlled by the tunnel coupling alone. Depending on whether the control parameter, the detuning and/or tunnel coupling, is changed adiabatically or diabatically, a C-phase or exchange gate, or a mixture of the two, is obtained. In the following sections, the C-phase and exchange gates are analyzed.

\subsubsection*{The C-Phase Gate}
In case the control parameter changes slowly, i.e.~adiabatically, the resulting operation can be described by the diagonal matrix (Appendix \ref{A:2qubit_cz}):
\begin{equation}
U_{cz,lab}(t) = \begin{bmatrix}
 e^{-i  t  \omega_{\lambda_1}} & 0 & 0 & 0 \\
 0 & e^{-i  t  \omega_{\lambda_2}} & 0 & 0 \\
 0 & 0 & e^{-i  t  \omega_{\lambda_3}} & 0 \\
 0 & 0 & 0 & e^{-i  t  \omega_{\lambda_4}} \\
\end{bmatrix},
\end{equation}
In the rotating frame this becomes:
\begin{equation}
\label{eq:2_u_cphase}
U_{cz,rot}(t) = \begin{bmatrix}
 1 & 0 & 0 & 0 \\
 0 & e^{-i  \phi_{Z,A}} & 0 & 0 \\
 0 & 0 & e^{-i \phi_{Z,B}} & 0 \\
 0 & 0 & 0 & 1 \\
\end{bmatrix},
\end{equation}
where $\phi_{Z,A} = t (\omega_{\lambda_2} - \frac{\delta\omega_{0}}{2})$ and $\phi_{Z,B} = t (\omega_{\lambda_3} + \frac{\delta\omega_{0}}{2})$. Two additional Z-rotations with angles $\phi_{Z,A}$ and $\phi_{Z,B}$ can be applied to the right and left qubit, respectively, to obtain the C-phase gate:
\begin{equation}
U_{cz}(t) = \begin{bmatrix}
 1 & 0 & 0 & 0 \\
 0 & 1 & 0 & 0 \\
 0 & 0 & 1 & 0 \\
 0 & 0 & 0 & e^{-i \theta_{cz}} \\
\end{bmatrix},
\end{equation}
where $\theta_{cz} = -(\phi_{Z,A} + \phi_{Z,B}) = -(\omega_{\lambda_2} + \omega_{\lambda_3}) t = \omega_{op} t$. These Z-rotations can easily be obtained by updating the software reference frame \cite{vandersypen2001experimental,vandersypen2005nmr}. In case $\theta_{cz} = \pi$, a controlled-Z operation is obtained.

Interestingly, the total acquired phase ($\phi_{Z,A} + \phi_{Z,B}$) is independent of $\delta\omega_{0}$. However, in case $\delta\omega_{0} = \sqrt{2} t_0$, $\phi_{Z,A} = \phi_{Z,B}$ \cite{coish2007exchange,meunier2011efficient,Veldhorst2015}, whereas for $\delta\omega_{0} = 0$, $\phi_{Z,A} = 0$.

\subsubsection*{The Exchange Gate}
If, instead, the control parameter is changed rapidly, i.e.~diabatically, and the Larmor frequency difference is negligible ($\delta\omega_0 \ll \omega_{op}$), the resulting operation follows as (Appendix \ref{A:2qubit_j}):
\begin{equation}
U_{J'}(t) \approx \begin{bmatrix}
 e^{-i t \omega_{\lambda_1}} & 0 & 0 & 0 \\
 0 & \frac{e^{-i t \omega_{\lambda_{2}}}+e^{-i t \omega_{\lambda_{3}}}}{2} & \frac{e^{-i t \omega_{\lambda_{2}}}-e^{-i t \omega_{\lambda_{3}}}}{2} & 0\\
 0 & \frac{e^{-i t \omega_{\lambda_{2}}}-e^{-i  t  \omega_{\lambda_{3}}}}{2} & \frac{e^{-i  t \omega_{\lambda_{2}}}+e^{-i  t  \omega_{\lambda_{3}}}}{2} & 0\\
 0 & 0 & 0 & e^{-i  t  \omega_{\lambda_4}} \\
\end{bmatrix}.
\end{equation}
In the rotating frame this becomes the exchange gate:
\begin{equation}
\label{eq:2_u_exchange}
U_{J}(t) \approx \begin{bmatrix}
 1 & 0 & 0 & 0 \\
 0 & \frac{1+e^{i \theta_{J}}}{2} & \frac{1-e^{i \theta_{J}}}{2} & 0\\
 0 & \frac{1-e^{i \theta_{J}}}{2} & \frac{1+e^{i \theta_{J}}}{2} & 0\\
 0 & 0 & 0 & 1 \\
\end{bmatrix},
\end{equation}
where $\theta_{J} = -\omega_{\lambda_{3}} t = \omega_{op} t$. In case $\theta_{J} = \pi$, a SWAP operation is obtained.

Note that since for an accurate exchange operation the Larmor frequency difference should be sufficiently small, the possibility of using FDMA for single-qubit operations (Section \ref{sec:1q_fdma}) is limited.
\newline

From Eq.\,\ref{eq:2_ham} it follows that the 2-qubit operations are affected by the Larmor frequencies ($\omega_{0,A},\omega_{0,B}$), the tunnel coupling ($t_0$), charging energy ($U$) and detuning ($\epsilon$). Furthermore, the operation depends on the total duration ($T$) that the 2-qubit gate is active. The effect of errors, both static and dynamic, on the fully electrically controlled parameters ($t_0$, $\epsilon$ and $T$) is analyzed in the subsequent section. Detailed derivations of the formulas can be found in Appendix \ref{A:2qubit}.

\subsection{Specifications for an exchange and C-phase gate}
In case of the C-phase gate, the control parameter is changed adiabatically, and the overall 2-qubit operation $U_{cz} = U_{out} \cdot U_{op} \cdot U_{in}$ consists of three parts: $U_{in}$ describing the adiabatic change to the operating point, $U_{op}$ describing the operation at the operating point and $U_{out}$ describes the adiabatic change to a point where the interaction is turned off sufficiently. As long as the adiabatic parts are not too slow, most of the operation occurs at the desired operating point as the exchange interaction is strongest there (Eq. \ref{eq:2_speed}). Consequently, only the effects of signal inaccuracy and noise on $U_{op}$ will be considered. Since for the exchange gate the control parameter changes diabatically, again the analysis will be limited to $U_{op}$ only.

\begin{table*}
\centering
\caption{The fidelity in case of control signal inaccuracies for various two-qubit operations.}
\label{tab:2_fidelities}
\begin{ruledtabular}
\begin{tabular}{l|llll}
 & \multicolumn{1}{c}{\textbf{SWAP}} & \multicolumn{1}{c}{\textbf{CZ}} & \multicolumn{1}{c}{\textbf{CZ}} & \multicolumn{1}{c}{\textbf{CZ}} \\
 & \multicolumn{1}{c}{$\delta\omega_0 = 0$} & \multicolumn{1}{c}{$\delta\omega_0 = 0$} & \multicolumn{1}{c}{$\delta\omega_0 = \omega_{op}$} & \multicolumn{1}{c}{$\delta\omega_0 = \sqrt{2} t_0$} \\
 Rotation $|\phi_{Z,B}|$ &  & \multicolumn{1}{c}{$\theta_{cz}$} & \multicolumn{1}{c}{$\frac{\theta_{cz}}{\sqrt{2}}$} & \multicolumn{1}{c}{$\frac{\theta_{cz}}{2}$} \\
\hline
Duration & $1 - \frac{3}{16} \theta_{J}^2 \left(\frac{\Delta T}{T}\right)^2$ & $1 - \frac{3}{16} \theta_{cz}^2 \left(\frac{\Delta T}{T}\right)^2$ & $1 - \frac{7-4\sqrt{2}}{16} \theta_{cz}^2 \left(\frac{\Delta T}{T}\right)^2$ & $1 - \frac{1}{16} \theta_{cz}^2 \left(\frac{\Delta T}{T}\right)^2$ \\
Tunnel coupling & $1 - \frac{3}{4} \theta_{J}^2 \left(\frac{\Delta t_0}{t_0}\right)^2$ & $1 - \frac{3}{4} \theta_{cz}^2 \left(\frac{\Delta t_0}{t_0}\right)^2$ & $1 - \frac{1}{2} \theta_{cz}^2 \left(\frac{\Delta t_0}{t_0}\right)^2$ & $1 - \frac{1}{4} \theta_{cz}^2 \left(\frac{\Delta t_0}{t_0}\right)^2$ \\
Detuning ($|\epsilon| > 0$) & $1 - \frac{3}{4} \theta_{J}^2 
\left(\frac{\frac{\epsilon}{U}}{1-\left(\frac{\epsilon}{U}\right)^2}\right)^2 \left(\frac{\Delta \epsilon}{U}\right)^2$
& $1 - \frac{3}{4} \theta_{cz}^2 
\left(\frac{\frac{\epsilon}{U}}{1-\left(\frac{\epsilon}{U}\right)^2}\right)^2 \left(\frac{\Delta \epsilon}{U}\right)^2$ & $1 - \frac{1}{2} \theta_{cz}^2 
\left(\frac{\frac{\epsilon}{U}}{1-\left(\frac{\epsilon}{U}\right)^2}\right)^2 \left(\frac{\Delta \epsilon}{U}\right)^2$ & $1 - \frac{1}{4} \theta_{cz}^2 
\left(\frac{\frac{\epsilon}{U}}{1-\left(\frac{\epsilon}{U}\right)^2}\right)^2 \left(\frac{\Delta \epsilon}{U}\right)^2$ \\
Detuning ($\epsilon = 0$) & $1 - \frac{3}{16} \theta_{J}^2 \left(\frac{\Delta \epsilon}{U}\right)^4$ & $1 - \frac{3}{16} \theta_{cz}^2 \left(\frac{\Delta \epsilon}{U}\right)^4$ & $1 - \frac{1}{8} \theta_{cz}^2 \left(\frac{\Delta \epsilon}{U}\right)^4$ & $1 - \frac{1}{16} \theta_{cz}^2 \left(\frac{\Delta \epsilon}{U}\right)^4$
\end{tabular}
\end{ruledtabular}
\end{table*}

The resulting fidelity in case of control signal inaccuracies has been summarized in Table \ref{tab:2_fidelities} for the exchange gate and the C-phase gate, both at zero detuning and finite detuning. For the exchange gate, we assume that no Larmor frequency difference between the qubits exists, since for such a gate $\delta\omega_0 \ll \omega_{op}$ is required, while for the C-phase gate different scenario's have been analyzed ($\delta\omega_0 = 0$, $\delta\omega_0 = \omega_{op}$ and $\delta\omega_0 = \sqrt{2} t_0$). In this table, $T$ denotes the qubit gate operation time, and inaccuracies are shown with the prefix $\Delta$. The table provides values for different rotation angles ($\theta_{cz}$, $\theta_{J}$). In case of the C-phase gate, additional Z-rotations might be required which only have a very small effect on the fidelity (Eq.\,\ref{eq:1_acc_z}).

The error contributions have a quadratic relation with the infidelity except for detuning errors for $\epsilon=0$ where a fourth-order dependence is found. This implies an improved robustness to detuning errors when operating in the charge symmetry point ($\epsilon=0$) \cite{reed2016reduced,martins2016noise}. 
Taking as example a $\pi$-gate targeting a 99.9\,\% fidelity (either SWAP or CZ at $\delta\omega_0 = 0$), the differences in the required detuning accuracy can be significant:
$\Delta\epsilon/U < 0.023\,\%$ when operating at $\epsilon \approx 0.99U$,
$\Delta\epsilon/U < 0.25\,\%$ when operating at $\epsilon \approx 0.9U$ and only
$\Delta\epsilon/U < 15\,\%$ when operating at the charge symmetry point.
The required relative timing and tunnel coupling accuracy are however more independent of the operating point and are $\sim$2.3\,\% and $\sim$1.2\,\%, respectively.

For low-frequency variations, i.e.~changing over a time-scale longer than the operation time, the expected fidelity follows the same equations as given in Table \ref{tab:2_fidelities} when replacing the inaccuracy, such as $\Delta\epsilon$, with the standard deviation, such as $\sigma\epsilon$, of the variation (assuming a Gaussian distribution, see Appendix \ref{A:method}). An exception is for detuning errors when operating at the charge symmetry point, because of the 4$^{\text{th}}$ order dependence. For a static but random error $\Delta$ for which $F = 1-c\cdot\Delta^4$, the expected fidelity is $F = 1-3\cdot c \cdot \sigma^4$, if $\Delta$ follows a Gaussian distribution with standard deviation $\sigma$ and zero mean (see Appendix \ref{A:method}). Consequently, there is a slightly higher sensitivity to noise than to static errors.

For timing variations, only the total duration matters and high-frequency noise is filtered as described by Eq.\,\ref{eq:1_jitter}. Moreover, similar as for the single-qubit gate, numerical simulations of the Hamiltonian have shown a sensitivity to high-frequency noise ($> \omega_{op}$) only in a limited band set by the duration of the operation, for both the electrically controlled detuning energy and tunnel coupling in case of the two-qubit gate. The quantum state is however also affected by noise around the frequencies corresponding to allowed energy transitions, in a band set by the duration of the operation. 
Consequently, it is important that the high-frequency noise components in the signals applied to the barrier gates and plunger gates are properly filtered. However, since the exchange gate requires a diabatic change in the control parameter, only limited filtering can be applied.

\subsection{Specifications for the idle operation}
\label{sec:2q_nop}
Since in practice the tunnel coupling cannot be fully removed, the two-qubit operation is never completely turned off (Eqs.\,\ref{eq:2_speed}).
The interaction strength can, however, be slowed down significantly, thus leading to a fidelity with respect to the ideal identity operation for the exchange and C-phase gates of (Appendix \ref{A:2qubit_nop}):
\begin{equation}
F_I =
\begin{cases}
1 - \frac{3}{16} \cdot \omega_{op,off}^2 \cdot T_{nop}^2 & \delta\omega_0 = 0\\
1 - \frac{7-4\sqrt{2}}{16} \cdot \omega_{op,off}^2 \cdot T_{nop}^2 & \delta\omega_0 = \omega_{op}\\
1 - \frac{1}{16} \cdot \omega_{op,off}^2 \cdot T_{nop}^2 & \delta\omega_0 = \sqrt{2} t_0
\end{cases},
\end{equation}
%
where $\omega_{op,off}$ is the reduced interaction strength during the time $T_{nop}$ when no operation is applied.

For a 99.9\,\% fidelity of an identity that takes 10 times longer than the nominal $\pi$-gate ($\theta_{J} = \theta_{cz} = \pi$), the interaction strength should be reduced by a factor $\sim$430 for both the exchange and C-phase gates ($\delta\omega_0 = 0$). Following Eq.\,\ref{eq:2_speed}, this could be achieved by lowering the tunnel coupling by a factor $\sim$21 while not changing the detuning. A 2-qubit operation performed at finite detuning could also be controlled using only the detuning. Assuming the interaction is considered off at zero detuning, the operation should be performed at a detuning of at least 99.9\,\% of $U$, which is very close to the avoided crossing. As mentioned before, operating closer to the avoided crossing reduces the tolerance to inaccuracies and noise in the detuning (Table \ref{tab:2_fidelities}).

\subsection{Case study of the specifications for a two-qubit operation}
\label{sec:2q_specs}

\begin{table*}[htb]
  \centering
  \caption{Example specifications for the control electronics when operating a C-phase gate at zero detuning. The PSD values provided as a comment assume a white spectrum with an ENBW $\approx$ 10\,MHz.}
  \begin{ruledtabular}
    \begin{tabular}{lrllr}
          & \multicolumn{1}{l}{\textbf{Value}} & \multicolumn{2}{c}{\textbf{Infidelity contribution}} & \multicolumn{1}{l}{\textbf{Comment}} \\
          &       & \textbf{to an operation} & \textbf{to idling} &  \\
    \hline
    \textit{\textbf{Frequency}} &       &       &       & \multicolumn{1}{l}{Following Eq. \ref{eq:1_off_omega}} \\
    spacing & \multicolumn{1}{l}{1\,GHz} &       &       &  \\
    inaccuracy & \multicolumn{1}{l}{11\,kHz} & 77 $\times 10^{-6}$ & 308 $\times 10^{-6}$ &  \\
    oscillator noise & \multicolumn{1}{l}{11\,kHz$_{rms}$} & 77 $\times 10^{-6}$ & 308 $\times 10^{-6}$  & \\
    nuclear spin noise & \multicolumn{1}{l}{1.9\,kHz$_{rms}$} & 2.2 $\times 10^{-6}$ & 8.9 $\times 10^{-6}$ &  \\
    \hline
    \textit{\textbf{Charging energy}} &       &       &       &  \\
    nominal & \multicolumn{1}{l}{83\,mV (4.1\,meV, 1.0\,THz)} &       &       &  \\
    \hline
    \textit{\textbf{Duration}} &       &       &       &  \\
    nominal & \multicolumn{1}{l}{250\,ns} &       &       & \multicolumn{1}{l}{$\omega_{op}$ = 2\,MHz} \\
    error & \multicolumn{1}{l}{5.3\,ns} & 281 $\times 10^{-6}$ &       &  \\
    \hline
    \textit{\textbf{Detuning energy}} &       &       &       &  \\
    nominal & \multicolumn{1}{l}{0\,mV (0\,$\mu$eV, 0\,GHz)} &       &       &  \\
    error & \multicolumn{1}{l}{12\,mV (0.60\,meV, 0.15\,THz)} & 281 $\times 10^{-6}$ &       & \multicolumn{1}{l}{$\sigma$ = 9.2\,mV$_{rms}$, PSD = 2.9\,$\mu$V/$\sqrt{\text{Hz}}$} \\
    \hline
    \textit{\textbf{Tunnel coupling}} &       &       &       &  \\
    nominal & \multicolumn{1}{l}{0.71\,GHz (2.9\,$\mu$eV)} &       &       &  \\
    error & \multicolumn{1}{l}{7.5\,MHz (31\,neV)} & 281 $\times 10^{-6}$ &       &  \\
    off-value & \multicolumn{1}{l}{78\,MHz (0.32\,$\mu$eV)} &       & 374 $\times 10^{-6}$ &  \\
    \hline
     &       & \textbf{$F_{cz}$ = 99.9\,\%} & \textbf{$F_I$ = 99.9\,\%} &  \\
    \end{tabular}%
  \end{ruledtabular}
  \label{tab:specs_2a}%
\end{table*}

Specifications for the control electronics responsible for the two-qubit operation can now be derived. This example develops on the example given in Section \ref{sec:1q_specs} and, for instance, assumes the same oscillator is used to keep the coherence with the qubits. Two examples will be given here, one at zero detuning and one at finite detuning. Both focus on the C-phase gate, operating at $\delta\omega_0 = \sqrt{2} t_0$. This choice for the Larmor frequency difference gives the most relaxed specifications for the control electronics, while at smaller $\delta\omega_0$ the specifications can be up to $\sqrt{3}$ times more demanding (see Table \ref{tab:2_fidelities}).

The Larmor frequency difference was chosen as 1\,GHz, to achieve a two-qubit operation speed of $\omega_{op}$ = 2\,MHz at zero detuning, while maintaining $\delta\omega_0 = \sqrt{2}\cdot t_0$ (Eq.\,\ref{eq:2_speed}). Example specifications for this operation are given in Table \ref{tab:specs_2a}. To further increase the operating speed, an even higher qubit frequency spacing would be required or $\delta\omega_{0} < \sqrt{2}\cdot t_0$. Alternatively, the operating speed can be enhanced to e.g.~20\,MHz, by operating the C-phase gate at finite detuning (Eq.\,\ref{eq:2_speed}), as shown in another example (Table \ref{tab:specs_2b}).

Both examples target a fidelity of 99.9\,\% for a C-phase gate with $\theta_{cz} = \pi$. The examples also indicate the specifications required for idling two qubits at 99.9\,\% fidelity for a duration of 500\,ns, the same as for the example in Section \ref{sec:1q_specs}.

For the charging energy and tunnel coupling typical values have been chosen. As the relation to the gate voltage is device dependent, no values for the required electrical specifications are given. Note that in either example, the tunnel coupling only has to change by a factor $\sim$ 9 to turn the operation on or off. In case of operation at finite detuning, this assumes zero detuning is applied when the operation is turned off.

The detuning energy is directly related to the voltage on the plunger gate via the lever arm, for which a typical value of $\alpha$ = 0.05\,eV/V has been assumed \cite{Kawakami2014}. When operating at finite detuning, the detuning energy is chosen at 95\,\% of the charging energy. Even though higher operating speeds can be obtained by moving even closer to the avoided crossing, the electrical specifications become increasingly challenging. Table \ref{tab:specs_2a} indicates that when operating at the charge symmetry point, very large detuning errors can be tolerated (at which point approximations used to derive the expressions in Table \ref{tab:2_fidelities} do not hold anymore). When operating at moderate detuning (Table \ref{tab:specs_2b}), the error specification for the detuning is more than 100 times stricter. Moreover, as the operation at finite detuning is faster with the same tunnel coupling, the signal bandwidth must be larger, with a larger noise bandwidth. As a rough estimate, the ENBW has been chosen as 5 times the operating speed in both examples, which seems plausible as an adiabatic change is required (for the exchange gate the situation might be worse). As a result, the maximum allowed noise spectral density, assuming white noise, is much lower. For the given example, this results in almost 5 orders of magnitude difference in the noise power spectral density.

In the example of Table \ref{tab:specs_2b}, the detuning control can be achieved by an AWG running at a sample rate of 1\,GS/s for a maximum timing inaccuracy of 0.5\,ns. Assuming the AWG has to cover a voltage range of $-U$...$U$ ($U$ is the charging energy), it must have a 10-bit resolution to meet the accuracy specification of the detuning energy.


\begin{table*}[htb]
  \centering
  \caption{Example specifications for the control electronics when operating a C-phase gate at finite detuning. The PSD values provided as a comment assume a white spectrum with an ENBW $\approx$ 100\,MHz.}
  \begin{ruledtabular}
    \begin{tabular}{lrllr}
          & \multicolumn{1}{l}{\textbf{Value}} & \multicolumn{2}{c}{\textbf{Infidelity contribution}} & \multicolumn{1}{l}{\textbf{Comment}} \\
          &       & \textbf{to an operation} & \textbf{to idling} &  \\
    \hline
    \textit{\textbf{Frequency}} &       &       &       & \multicolumn{1}{l}{Following Eq. \ref{eq:1_off_omega}} \\
    spacing & \multicolumn{1}{l}{1\,GHz} &       &       &  \\
    inaccuracy & \multicolumn{1}{l}{11\,kHz} & 0.8 $\times 10^{-6}$ & 308 $\times 10^{-6}$ &  \\
    oscillator noise & \multicolumn{1}{l}{11\,kHz$_{rms}$} & 0.8 $\times 10^{-6}$ & 308 $\times 10^{-6}$  & \\
    nuclear spin noise & \multicolumn{1}{l}{1.9\,kHz$_{rms}$} & 0.02 $\times 10^{-6}$ & 8.9 $\times 10^{-6}$ &  \\
    \hline
    \textit{\textbf{Charging energy}} &       &       &       &  \\
    nominal & \multicolumn{1}{l}{83\,mV (4.1\,meV, 1.0\,THz)} &       &       &  \\
    \hline
    \textit{\textbf{Duration}} &       &       &       &  \\
    nominal & \multicolumn{1}{l}{25\,ns} &       &       & \multicolumn{1}{l}{$\omega_{op}$ = 20\,MHz} \\
    error & \multicolumn{1}{l}{0.58\,ns} & 333 $\times 10^{-6}$ &       &  \\
    \hline
    \textit{\textbf{Detuning energy}} &       &       &       &  \\
    nominal & \multicolumn{1}{l}{78\,mV (3.9\,meV, 0.95\,THz)} &       &       &  \\
    error & \multicolumn{1}{l}{0.10\,mV (5.1\,$\mu$eV, 1.2\,GHz)} & 333 $\times 10^{-6}$ &       & \multicolumn{1}{l}{$\sigma$ = 0.10\,mV$_{rms}$, PSD = 10\,nV/$\sqrt{\text{Hz}}$} \\
    \hline
    \textit{\textbf{Tunnel coupling}} &       &       &       &  \\
    nominal & \multicolumn{1}{l}{0.71\,GHz (2.9\,$\mu$eV)} &       &       &  \\
    error & \multicolumn{1}{l}{8.2\,MHz (34\,neV)} & 333 $\times 10^{-6}$ &       &  \\
    off-value & \multicolumn{1}{l}{78\,MHz (0.32\,$\mu$eV)} &       & 374 $\times 10^{-6}$ &  \\
    \hline
     &       & \textbf{$F_{cz}$ = 99.9\,\%} & \textbf{$F_I$ = 99.9\,\%} &  \\
    \end{tabular}%
  \end{ruledtabular}
  \label{tab:specs_2b}%
\end{table*}


\section{\label{sec:readout} Signal Specifications for Qubit Read-Out}
For the read-out of the quantum state, the Pauli spin-blockade read-out \cite{ono2002current} is analyzed since it offers several advantages with respect to the other possible alternative, i.e.~the Elzerman read-out \cite{elzerman2004single}: no electron reservoir is required next to the quantum dot; the Zeeman energy splitting does not have to be much higher than the thermal energy, thus enabling operation at higher temperatures and/or lower Larmor frequencies. As a drawback, the Pauli spin-blockade read-out involves two quantum dots, where the measurement involves the discrimination between the singlet and triplet states.

Even though relaxation, which is quantified with the relaxation time $T_1$, is an important limiting factor in qubit read-out, its effect is not considered in the following analysis as it is not influenced by the control electronics. Furthermore, in our analysis, we assume that the spin-dependent charge state resulting from a Pauli spin-blockade read-out is measured using a charge sensor. 
As a result, the read-out fidelity is determined by various factors:
%
\begin{itemize}
    \item $P_{charge}$: the probability that the spin-state is correctly projected to the charge state.
    \item $P_{sense}$: the probability that the charge sensor correctly detects the charge state.
    \item $P_{detect}$: the probability that the read-out circuit correctly discriminates the signal of the charge sensor.
\end{itemize}
The overall read-out fidelity then follows as:
\begin{equation}
\label{eq:qr_fidelity_ps}
F \approx P_{charge} \cdot P_{sense} \cdot P_{detect}.
\end{equation}
%
The probability $P_{sense}$ is limited by e.g.~
interference on one of the charge-sensor bias gates and charge noise in the substrate. As this highly depends on the type of sensor employed and the sensor integration, this error contribution will not be discussed further.

The quantum-dot control electronics limit $P_{charge}$, as discussed in Section \ref{sec:readout_control}, while the read-out electronics limit $P_{detect}$, as discussed in Section \ref{sec:readout_readout}.

\subsection{Specifications for the electronics controlling the spin to charge conversion}
\label{sec:readout_control}
\begin{figure}
\includegraphics[width=\linewidth]{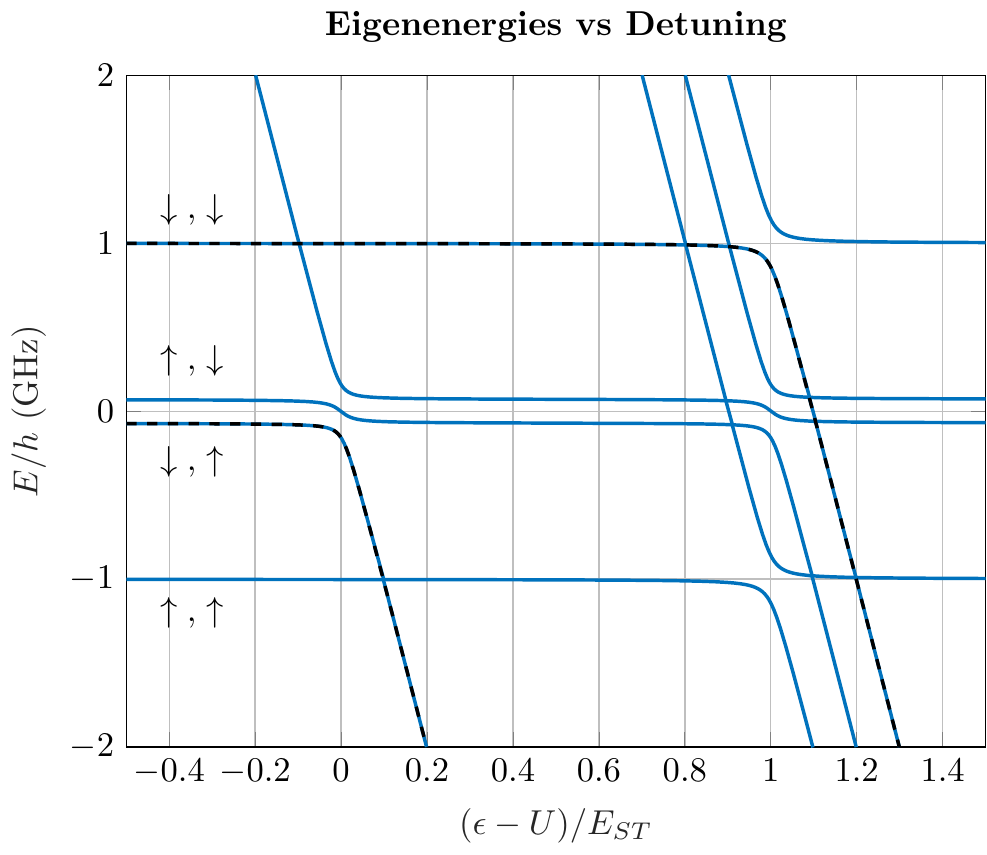}
\caption{\label{fig_qr_energies} The energy of the stationary states versus the detuning near the avoided crossing. The black dashed lines indicate the states where the left qubit was originally in the $\left| \downarrow \right\rangle$ state.}
\end{figure}

For the analysis of the charge transfer in the Pauli spin-blockade read-out, the Hamiltonian of Eq.\,\ref{eq:2_ham} is extended with the lowest-energy triplet states (either due to the valley splitting or the orbital energy splitting). Those states are spaced by a singlet-triplet energy splitting $E_{ST}$ from the singlet energy level (see Appendix \ref{A:readout} for the Hamiltonian). A plot of the energy of the stationary states versus the detuning near the avoided crossing is shown in Fig.\,\ref{fig_qr_energies}. For the following discussion, only the $\left|\downarrow,\downarrow\right\rangle$ and $\left|\downarrow,\uparrow\right\rangle$ states, highlighted in Fig.\,\ref{fig_qr_energies}, need to be considered.

The Pauli spin-blockade read-out relies on $E_{ST}$ for the discrimination of the single-dot singlet configuration from three possible single-dot triplet configurations.
Considering a pair of neighboring qubits, the state of the right qubit can be measured as follows. The left qubit is initialized in the $\left|\downarrow\right\rangle$ state.
By detuning adiabatically to a point between the singlet avoided crossing and the triplet avoided crossing (with $t_0 > 0$), only the $\left|\downarrow,\uparrow\right\rangle$-state (at $\epsilon = 0$) becomes a singlet and both electrons will move into the same dot. This charge movement can be measured using a charge sensor. Based on the measurement result, it is then clear whether the qubits are in a singlet or one of the three triplet configurations. 
This scenario will be analyzed here.

Starting from the $\left|\downarrow,\downarrow\right\rangle$-state, there is a small probability $P(\text{transfer}|\left|\downarrow,\downarrow\right\rangle)$ that both electrons end up in the same dot. Similarly, starting from the $\left|\downarrow,\uparrow\right\rangle$-state, there is a small probability $P(\text{no transfer}|\left|\downarrow,\uparrow\right\rangle)$ that no charge will transfer.
The probability of a correct spin to charge conversion can be defined as:
%
\begin{equation}
P_{charge} = 1 - P(\text{transfer}|\left|\downarrow,\downarrow\right\rangle) - P(\text{no transfer}|\left|\downarrow,\uparrow\right\rangle).
\end{equation}
The analysis is again simplified by assuming an ideal adiabatic change in the detuning energy. 
The results presented in this section are obtained from numerical simulations of the Hamiltonian.

\begin{figure}
\includegraphics[width=\linewidth]{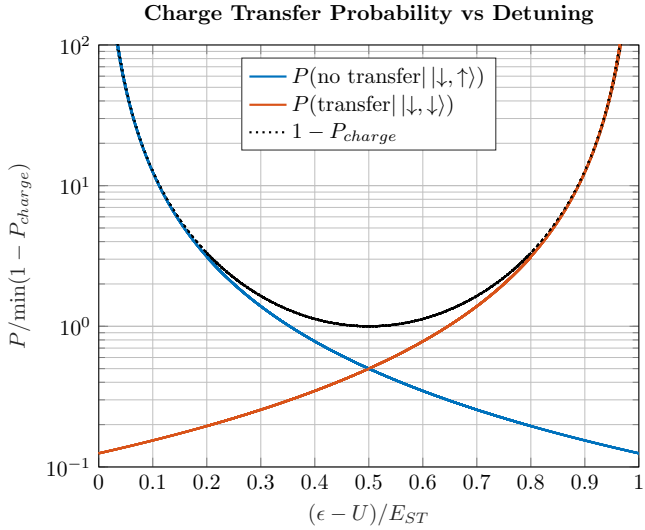}
\caption{\label{fig_qr_shape} The probability $P_{charge}$ at various points of detuning as simulated for various tunnel rates, Larmor frequencies, charging energies and singlet-triplet energy splittings (each varied over a decade; the resulting plots are overlapping). The obtained probabilities are plotted relative to the minimum probability that was obtained in each case (i.e.~the probability at $\epsilon = U + \frac{E_{ST}}{2}$).}
\end{figure}

\begin{figure}[htbp]
\includegraphics[width=\linewidth]{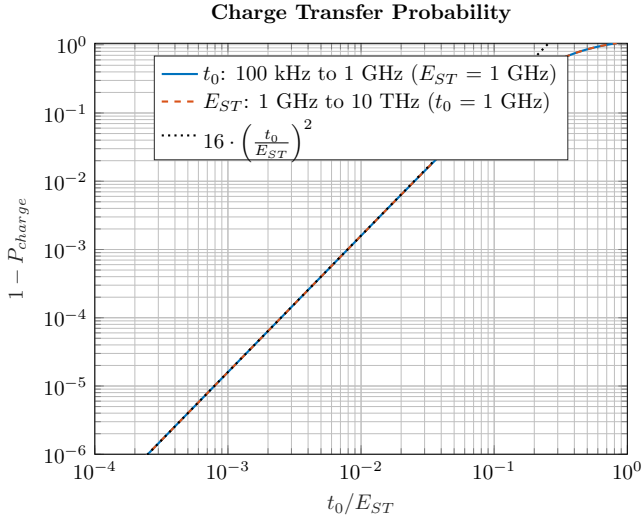}
\caption{\label{fig_qr_fidelity} The simulated probability $1-P_{charge}$ versus the singlet-triplet splitting, normalized to the tunnel coupling at $\epsilon = U + \frac{E_{ST}}{2}$, while sweeping either the tunnel coupling or the singlet-triplet energy splitting}
\end{figure}

Simulations show that the highest $P_{charge}$ is obtained by detuning to $\epsilon = U + \frac{E_{ST}}{2}$, i.e.~equidistant between the  singlet and triplet avoided crossings, as shown in Fig.\,\ref{fig_qr_shape}. The shape of the probability versus $\frac{\epsilon-U}{E_{ST}}$ plot is independent of the Larmor frequency, assuming $\omega_0 \ll E_{ST}$. Although the shape remains the same, the obtainable maximum $P_{charge}$ scales with tunnel coupling and singlet-triplet energy splitting, as can be seen in Fig.\,\ref{fig_qr_fidelity}. From this figure, an upper bound for the tunnel coupling can be found, which must be maintained even with errors caused by limitations in the control electronics.

%
%
%

Even though $P_{charge}$ is highly influenced by the achievable tunnel couplings and singlet-triplet energy splittings in the system (Fig.\,\ref{fig_qr_fidelity}), the detuning value has a minor influence (provided there is a sufficient singlet-triplet energy splitting), since $1-P_{charge}$ is relatively flat around its minimum, as shown in Fig.\,\ref{fig_qr_shape}. For instance, for a twofold increase in $1-P_{charge}$, the detuning must stay in the range $\frac{\epsilon-U}{E_{ST}} \approx 0.5 \pm 0.235$. We can then conclude that a large singlet-triplet splitting is desired to limit the influence of the control electronics on the read-out.

\subsection{Specifications for the electronics processing the read-out signal}
\label{sec:readout_readout}

\begin{figure}
\includegraphics[width=\linewidth]{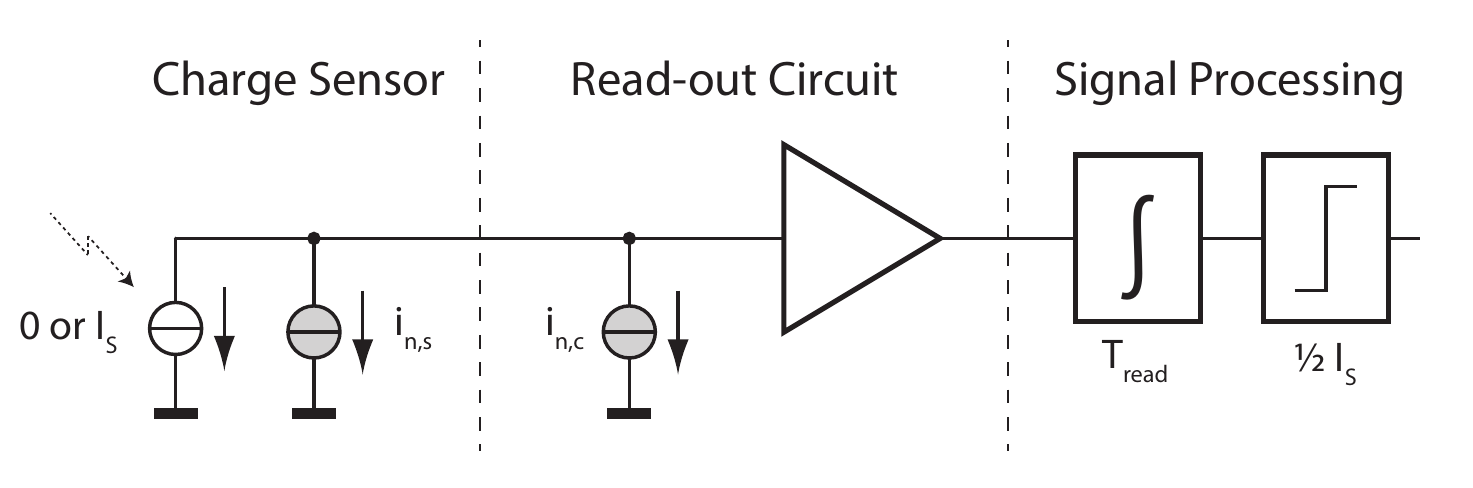}
\caption{\label{fig_qr_model} A model of a typical read-out chain, showing the sensor, the read-out electronics and the required signal processing for the measurement discrimination. Additional sources modeling the noise are shown in gray.}
\end{figure}

\begin{figure}
\includegraphics[width=\linewidth]{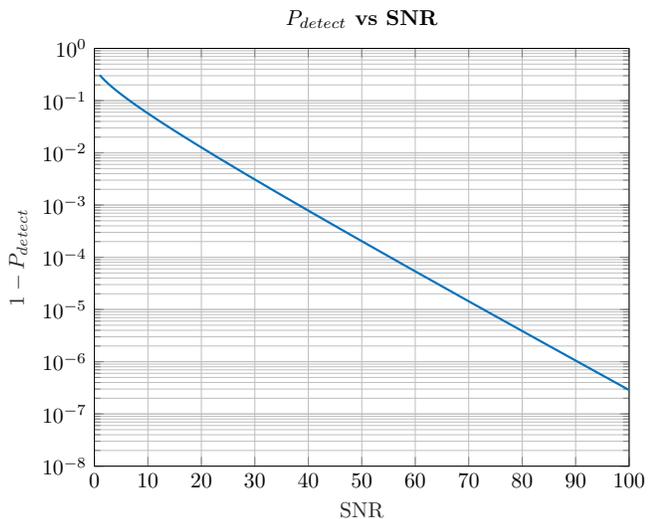}
\caption{\label{fig_qr_gaussian} A plot of $1-P_{detect}$ versus SNR in case of Gaussian distributed noise.}
\end{figure}

In this section, we will consider a direct read-out. A model of a typical read-out chain is shown in Fig.\,\ref{fig_qr_model}. For simplicity, the sensor will be modelled as a current source with a value of either 0 or $I_{s}$, depending on the sensed charge. The read-out fidelity is limited by the noise introduced by the sensor and by the read-out circuit, indicated in Fig.\,\ref{fig_qr_model} as $i_{n,s}$ and $i_{n,c}$, respectively. Assuming the typical matched-filter detection, i.e.~integrating the signal current for a duration $T_{read}$ and comparing the result to a threshold, the probability of a correct measurement under the presence of Gaussian-distributed noise is given by (see Appendix \ref{A:readout}):
\begin{equation}
P_{detect} = \frac{1 + \text{erf} \left(\sqrt{\frac{\text{SNR}}{8}} \right)}{2},
\end{equation}
with
\begin{equation}
\text{SNR} = \frac{I_s^2}{\int_0^\infty S_i(f) \cdot \left(\frac{\sin(\pi f T_{read})}{\pi f}\right)^2 df},
\end{equation}
where $S_i(f)$ is the PSD of the total noise $i_n = i_{n,s} + i_{n,c}$. In case the noise is white, this simplifies to:
\begin{equation}
\text{SNR} = \frac{I_s^2}{S_i \cdot ENBW},
\end{equation}
with effective noise bandwidth $ENBW = 1/(2 \cdot T_{read})$. The plot of $P_{detect}$ versus SNR (Fig.\,\ref{fig_qr_gaussian}) indicates that the signal ($I_s$) should be $\sim6$ times larger than the standard deviation of the noise ($\sqrt{S_i \cdot ENBW}$) for a 99.9\,\% probability of correct measurement discrimination.

\subsection{Case study of the specifications for a qubit read-out}
\label{sec:ro_specs}

\begin{table*}
  \centering
  \caption{Example specifications for the control electronics. The PSD values provided as a comment assume a white spectrum with an ENBW $\approx$ 1\,MHz for the detuning control, and a measurement time of $T_{read}$ = 0.6\,$\mu$s.}
  \begin{ruledtabular}
    \begin{tabular}{lrlr}
          & \multicolumn{1}{l}{\textbf{Value}} & \multicolumn{1}{c}{\textbf{Infidelity contribution}} & \multicolumn{1}{l}{\textbf{Comment}} \\
          &       & \textbf{to the read-out} &  \\
    \hline
    \textit{\textbf{Charging energy}} &       &       &  \\
    nominal & \multicolumn{1}{l}{82.7\,mV (4.1\,meV, 1.0\,THz)} &       &  \\
    \hline
    \textit{\textbf{Singlet-triplet energy}} &       &       &  \\
    nominal & \multicolumn{1}{l}{1.0\,mV (50\,$\mu$eV, 12\,GHz)} &       &  \\
    \hline
    \textit{\textbf{Detuning energy}} &       &       &  \\
    nominal & \multicolumn{1}{l}{83.2\,mV (4.2\,meV, 1.0\,THz)} &       &  \\
    error & \multicolumn{1}{l}{0.24\,mV (12\,$\mu$eV, 2.8\,GHz)} & 167 $\times 10^{-6}$ & \multicolumn{1}{l}{$\sigma$ = 0.24\,mV$_{rms}$, PSD = 0.24\,$\mu$V/$\sqrt{\text{Hz}}$} \\
    \hline
    \textit{\textbf{Tunnel coupling}} &       &       &  \\
    nominal & \multicolumn{1}{l}{39\,MHz (0.16\,$\mu$eV)} & 167 $\times 10^{-6}$ &  \\
    \hline
    \textbf{Contribution $P_{charge}$} &       & \textbf{99.967\,\%} &  \\
    \\
    \hline
    \textbf{Contribution $P_{sense}$}  &       & \textbf{99.967\,\%} &  \\
    \\
    \hline
    \textit{\textbf{Quantum Point Contact}} &       &       &  \\
    signal & \multicolumn{1}{l}{400\,pA} &       &  \\
    noise & \multicolumn{1}{l}{53\,pA$_{rms}$} &    222 $\times 10^{-6}$    & \multicolumn{1}{l}{PSD = 57\,fA$/\sqrt{\text{Hz}}$} \\
    \hline
    \textit{\textbf{Readout Circuit}} &       &       &  \\
    input-referred noise & \multicolumn{1}{l}{26\,pA$_{rms}$} &  111 $\times 10^{-6}$    & \multicolumn{1}{l}{PSD = 28\,fA$/\sqrt{\text{Hz}}$} \\
    \hline
    \textbf{Contribution $P_{detect}$} &       & \textbf{99.967\,\%} & \multicolumn{1}{l}{SNR = 46} \\
    \end{tabular}%
    \end{ruledtabular}
  \label{tab:specs_ro}%
\end{table*}

The example specifications presented in this section build on those presented in Section \ref{sec:1q_specs} and \ref{sec:2q_specs}, and hence assume the same charging energy and lever arm for the detuning control. For the singlet-triplet energy splitting a typical value of $E_{ST}$ = 50\,$\mu$eV has been used. As a result, the optimum detuning is at 83.2\,mV. The resulting specifications have been summarized in Table \ref{tab:specs_ro} and assume equal contributions from $P_{charge}$, $P_{sense}$ and $P_{detect}$ (Eq.\,\ref{eq:qr_fidelity_ps}).

Following Fig.\,\ref{fig_qr_fidelity}, to achieve the required $P_{charge}$, the tunnel coupling must be even lower than what was required to turn off the two-qubit operation (see Section \ref{sec:2q_specs}), thereby extending the required tunnel coupling tuning range to $\sim$ 18$\times$. In the example of Table \ref{tab:specs_ro}, the detuning control can be achieved by an AWG with low sample rate, as the detuning must change adiabatically and the read-out generally takes a relatively long time. Assuming the AWG has to cover a voltage range from 0 to 2$U$, it must have a 9-bit resolution to meet the accuracy specification of the detuning energy (Fig.\,\ref{fig_qr_shape}). As a result, the same circuitry used for the two-qubit operation (Section \ref{sec:2q_specs}) could potentially be used.

In this example, $P_{detect}$ assumes a direct read-out of a QPC, following the numbers provided in \cite{vink2007cryogenic} ($I_s$ = 400\,pA and $i_{n,s}$ = 57\,fA$/\sqrt{\text{Hz}}$). Assuming the read-out circuit is designed to contribute about half the noise compared to the shot noise limit of the QPC ($i_{n,c} \approx i_{n,s}/2$), an integration time of at least $T_{read}$ = 0.6\,$\mu$s is required to achieve an SNR of 46 for a $P_{detect}$ of 99.967\% (for F = 99.9\%, Eq. \ref{eq:qr_fidelity_ps}). For such short read-out times, the assumption of white noise is valid, and the effects of qubit relaxation ($T_1$) could be negligible.

%
%


\section{\label{sec:discussion} Discussion}
The electronics specifications derived in the examples outlined in Sections \ref{sec:1q_specs}, \ref{sec:2q_specs} and \ref{sec:ro_specs} for a target  process fidelity of 99.9\,\% will now be compared to the performance achieved by state-of-the-art electronics. To this purpose, table \ref{tab:instruments} summarizes the performance of commonly used arbitrary waveform generators and microwave vector sources \cite{Tektronix,Keysight2014_M9330A,Tabor,Keysight2016,RS2018}.


\begin{table*}[htbp]
  \centering
  \caption{Specifications of commonly used arbitrary waveform generators and microwave vector sources.}
  \begin{ruledtabular}
    \begin{tabular}{l|lllll}
          \multicolumn{1}{p{12em}}{} & \multicolumn{1}{p{7em}}{\textbf{Sample Rate}} & \multicolumn{1}{p{7em}}{\textbf{Resolution}} & \multicolumn{1}{p{7em}}{\textbf{Jitter}} & \multicolumn{1}{p{7em}}{\textbf{Output noise}} & \multicolumn{1}{l}{\textbf{SFDR}} \\
          \hline
    \textbf{Tektronix AWG5014C \cite{Tektronix}} & 1.2\,GS/s & 14\,bit & 5.0\,ps$_{rms}$ & -\footnotemark[1] & $<$ -56\,dBc \\
    \textbf{Keysight M9330A \cite{Keysight2014_M9330A}} & 1.25\,GS/s & 15\,bit & -     & -150\,dBc/Hz & $<$ -65\,dBc \\
    \textbf{Tabor WX1282C \cite{Tabor}} & 1.25\,GS/s & 14\,bit & -     & -     & $<$ -44\,dBc \\
    \end{tabular}%
  \end{ruledtabular}
  \footnotetext[1]{Amplitude resolution: 1 mV}
  \begin{tabular}{l}
  \\
  \end{tabular}
  \begin{ruledtabular}
    \begin{tabular}{l|llll}
           \multicolumn{1}{p{12em}}{} & \multicolumn{1}{p{7em}}{\textbf{Max. Output Frequency}} & \multicolumn{1}{p{7em}}{\textbf{Frequency Resolution}} & \multicolumn{1}{p{7em}}{\textbf{Phase Noise (100 kHz)}} & \multicolumn{1}{l}{\textbf{Wideband noise (10 dBm)}} \\
          \hline
    \textbf{Agilent E8267D \cite{Keysight2016}} & 44\,GHz & 1\,mHz & $<$ -100\,dBc/Hz & $<$ -141\,dBc/Hz \\
    \textbf{R\&S SMW200A \cite{RS2018}} & 40\,GHz & 1\,mHz & $<$ -100\,dBc/Hz & $<$ -134\,dBc/Hz \\
    \end{tabular}%
  \end{ruledtabular}
  \label{tab:instruments}%
\end{table*}

For the generation of the detuning control and the microwave envelope, an AWG is required. We will compare the specifications of the Tektronix 5014C, as used in e.g.~\cite{Kawakami2014,Kawakami2016,Watson2017}, with the specifications derived in the case study. This AWG achieves a sample rate of 1.2\,GS/s with 14-bit resolution, thereby providing enough resolution for the amplitude and duration of the microwave envelope (150\,MS/s and 8\,bits, see example in Section \ref{sec:1q_specs}). The worst-case spurious-free dynamic range of $-56$\,dBc is well below the required $-41$\,dBc. 
The specified random jitter of 5.0\,ps$_{rms}$ is well below the required value of 3.6\,ns$_{rms}$. Finally, the output noise level is not clearly specified, but can be assumed to be not much larger than the amplitude resolution of 1\,mV. In case the AWG's output is attenuated by $\sim$ 40\,dB, this also meets the specifications. This AWG can also be used for detuning control in two-qubit gates (example in Section \ref{sec:2q_specs}). The sample rate is high enough to meet the required timing resolution ($>$ 1\,GS/s), and the resolution is sufficiently high to reach the detuning requirements (worst-case 0.10\,mV for a $<$ 100\,mV pulse) with a 20\,dB attenuator. As the specifications for Pauli-spin blockade read-out (example in Section \ref{sec:ro_specs}) are more relaxed, the same AWG again suffices.

Finally, for the generation of the microwave carrier, some setups use the Agilent MW vector source E8267D \cite{Kawakami2014,Kawakami2016,Watson2017,yoneda201799,Veldhorst2014,Veldhorst2015}, which has a frequency resolution well below the requirements (1\,mHz vs $\sim$ 20\,kHz). The single-sided phase noise is also well below the required -106\,dBc/Hz at a 1\,MHz offset from the carrier (at the worst point the E8267D achieves better than -100\,dBc/Hz at a 100\,kHz offset). The broadband noise is specified as 63\,nV/$\sqrt{\text{Hz}}$ (-141\,dBc/Hz at 10\,dBm) and therefore at least 20\,dB attenuation is required to meet the specification of 7.1\,nV/$\sqrt{\text{Hz}}$.

It can be concluded that these instruments are capable of supporting a 99.9\% fidelity. However, these instruments are bulky, consume several Watts of power and can not be operated at cryogenic temperatures, therefore hindering scalability.

Fully integrated CMOS circuits operating at cryogenic temperatures can be adopted to tackle this problem \cite{Hornibrook2015, conway2016fpga,Homulle2016, Charbon2016, charbon2017cryocmos, patra2017cryocmos, Sebastiano2017}. In order to assess the feasibility of such solution, the power consumption of the required circuit blocks will be estimated by using room-temperature CMOS circuits as reference. This is valid, since cryogenic CMOS circuits are expected to show significantly less noise for the same power budget, as shown in \cite{charbon2017cryocmos, patra2017cryocmos}. As a result, the estimates given here likely overestimate the required power consumption. Furthermore, we will assume a 50-$\Omega$ load for each circuit, which is not the case  for a fully integrated controller.

The core component determining the specifications of an AWG is its digital-to-analog converter (DAC). The 10-bit 500\,MS/s DAC presented in \cite{Lin2012} meets the specifications for the microwave envelope generation at a power consumption of 24\,mW. For the detuning control, the DAC specifications are stricter, but can be met by the 12-bit 1.6\,GS/s DAC presented in \cite{Lin2014}, with a power consumption of 40\,mW. Although for the tunnel barrier the specifications will depend highly on the gate structure, a similar DAC is assumed to be sufficient.

The core component of a microwave carrier generator, the phase-locked loop (PLL), is also available as a CMOS circuit operating over the required frequency range (9.2 - 12.7\,GHz) at a power consumption around 13\,mW \cite{Raczkowski2015}. Its phase noise performance is slightly worse than required, achieving $-105$\,dBc/Hz at a 2\,MHz offset. However, with operation at cryogenic temperatures the noise level is expected to improve.

In a linear qubit array, one DAC is required for the barrier gate and one for the plunger gate for each qubit. This leads to an estimated power of 80\,mW per qubit. For the microwave signals, the envelope DAC and PLL together consume $\sim$ 40\,mW. Without any form of multiplexing, this indicates a power consumption of 120\,mW/qubit. For a state-of-the-art dilution refrigerator with a cooling power of a few Watts at 4\,K, this suggests a maximum of a few tens of qubits when operating the classical controller at 4\,K. 

However, the power consumption of the DACs controlling the barrier gates and plunger gates could be highly reduced if not being limited to a 50-$\Omega$ system. To get a sufficient signal swing, in \cite{Lin2014} a 16-mA current is delivered to a 50-$\Omega$ load, thereby setting a lower bound to the power consumption. A much lower current would be required for a higher impedance, or even for a lower swing as acceptable in this application, ultimately limited by the speed or noise requirement. Furthermore, the same fast DAC can be used to generate frequency multiplexed microwave envelopes. With a sample rate of 1.6\,GS/s, a bandwidth of roughly 640\,MHz is available \cite{Lin2014}. This can be used to drive 64\,qubits with a Rabi freqency of 1\,MHz spaced by 10\,MHz  using e.g.~a Gaussian envelope (Fig.\,\ref{fig_q1_fdma_2}). The combined power of the fast DAC and PLL, i.e.~53\,mW, is then shared over 64\,qubits, thus resulting in a power consumption below 1\,mW/qubit. For the read-out on the other hand, cryogenic CMOS circuits have already been proposed that can achieve a power consumption $<$ 1\,mW/qubit \cite{charbon2017cryocmos, patra2017cryocmos}.

In summary, a cryogenic CMOS controller for a large-scale quantum processor appears to be feasible for a target fidelity of 99.9\%. However, for a minimum power consumption, the trade-offs in the electronics design must be systematically investigated. The analysis proposed in this paper provides the foundations for such optimization and will help electronics designers to build a functional controller.

\section{\label{sec:conclusion} Conclusion}
In this paper, the effect of non-idealities in the classical controller for a quantum processor has been analyzed. Even though this work focused on single-electron spin-qubits, the presented approach can be used to analyze the performance of a quantum processor in any qubit technology. A comprehensive approach has been proposed, by covering the effect of both static and dynamic errors on all quantum operations, i.e. single-qubit gates, two-qubit gates and read-out.

With the results of this analysis, the impact of the controller on the performance of the quantum computer as a whole can be quantified. This is required to ensure that the controller does not become the performance bottleneck as the qubit performance keeps improving. Moreover, with the presented results, a full set of electrical specifications can be derived, targeting a given qubit fidelity, as exemplified in a case study targeting a 99.9\% fidelity for the various quantum operations. The availability of these specifications enables the design of next generation controllers tailor-made to the quantum processor, optimized for performance, power, cost and size, so as to improve the scalability of the quantum computer.

As future controllers might have to operate physically close to the quantum processor, i.e.~at cryogenic temperatures where the cooling power is limited, the power optimization of the controller will be critical in enabling large-scale quantum computing. With the results obtained in this paper, the trade-offs between qubit fidelity and power spent in the controller can be analyzed, this representing the foundation for such a power optimization.

\begin{acknowledgments}
The authors would like to thank Intel Corp. for funding, and Dr. Andrea Corna and Richard Versluis for the many interesting discussions and their useful insights.
\end{acknowledgments}


\appendix
\section{General Derivations}
\label{A:method}
The average fidelity of a unitary operation on an \mbox{n-dimensional} complex Hilbert space can be calculated as \cite{PEDERSEN2007}:
\begin{equation}
F_{av} = \frac{n + \left|\text{Tr}\left[U_{ideal}^\dag U_{real}\right]\right|^2}{n\cdot(n+1)}.
\end{equation}
Using the relation between the average gate fidelity ($F_{av}$) and process fidelity ($F$) \cite{nielsen2002simple}:
\begin{equation}
F_{av} = \frac{1 + n \cdot F}{n + 1},
\end{equation}
the process fidelity can be obtained from the unitary operations as:
\begin{equation}
\label{eq:app_fidelity_definition}
F = \frac{1}{n^2}\cdot\left|\text{Tr}\left[U_{ideal}^\dag U_{real}\right]\right|^2,
\end{equation}
which will be used throughout this paper, in line with \cite{green2012high,Green2013, Ball2016}. In case of random variations of the unitary operation $U_{real}$, the expected value of the fidelity will be evaluated:
\begin{equation}
\label{eq:app_fidelity_integral}
\langle F \rangle = \int_{-\infty}^{\infty} F(x) f(x) dx,
\end{equation}
where $F(x)$ is the fidelity for a random parameter $x$, and $f(x)$ is the probability density function of $x$.

In case of Gaussian distributed noise with zero mean and standard deviation $\sigma$:
\begin{equation}
f(x) = \frac{e^{-\frac{x^2}{2\sigma^2}}}{\sqrt{2\pi\sigma^2}}.
\end{equation}
For this distribution, and certain expressions for $F(x)$, the integral in Eq. \ref{eq:app_fidelity_integral} has a simple solution.

In case $F(x) = 1 - c \cdot x^2$:
\begin{eqnarray}
\langle F \rangle &=& \int_{-\infty}^{\infty} \left(1 - c \cdot x^2\right) \frac{e^{-\frac{x^2}{2\sigma^2}}}{\sqrt{2\pi\sigma^2}} dx
\\
\label{eq:app_gauss_2}
&=&
1 - c \cdot \sigma^2.
\end{eqnarray}

In case $F(x) = 1 - c \cdot x^4$:
\begin{eqnarray}
\langle F \rangle &=& \int_{-\infty}^{\infty} \left(1 - c \cdot x^4\right) \frac{e^{-\frac{x^2}{2\sigma^2}}}{\sqrt{2\pi\sigma^2}} dx
\\
&=&
1 - 3 \cdot c \cdot \sigma^4.
\end{eqnarray}

\subsection{Numerical Simulations}
The numerical simulations performed in case of microwave pulses with Gaussian envelopes use finite time steps to approximate the unitary operation:
\begin{equation}
U_{real} \approx \prod_{n=N}^0  e^{-i \cdot H(i \cdot \Delta t) \cdot \Delta t},
\end{equation}
where $H(t)$ is the Hamiltonian at time $t$. The time step $\Delta t$ is chosen to be constant and sufficiently small in order not to affect the simulation results. A 10 times oversampling with respect to the signal's carrier frequency was found to give accurate results. 

\section{Derivations for Single-Qubit Operation}
\label{A:1qubit}
Starting from the Hamiltonian that describes a single electron in the lab frame under microwave excitation ($\hbar = 1$)
\begin{equation}
\label{eq:app_ham_lab}
H_{lab} = -\omega_0 \cdot \frac{\sigma_z}{2} + \omega_{ESR}(t) \cdot \frac{\sigma_x}{2},
\end{equation}
where $\sigma_x$, $\sigma_y$ and $\sigma_z$ are the Pauli operators. For a microwave signal $\omega_{ESR}(t) = 2 \cdot \omega_R \cdot \cos(\omega_{mw} t + \phi)$, the Hamiltonian can be made time-independent by moving to a reference frame that rotates with a frequency $\omega_{mw}$ around the z-axis:
\begin{eqnarray}
\label{eq:app_rwa_1}
H_{ref} &=& -\omega_{mw} \cdot \frac{\sigma_z}{2}
\\
\label{eq:app_rwa_2}
U_{ref}(t) &=& e^{-i \cdot H_{ref} \cdot t}.
\end{eqnarray}
The Hamiltonian in the rotating frame follows as:
\begin{eqnarray}
\label{eq:app_rwa_3}
&&H_R(t) = U_{ref}(t)^\dag \cdot \left(H_{lab} - H_{ref}\right) \cdot U_{ref}(t)
\\
&=& \frac{\omega_R}{2} \begin{bmatrix}  
 \frac{\omega_{mw} - \omega_0}{\omega_R} & e^{i\phi}+e^{-i\phi}e^{-2i\omega_{mw}t}\\
 e^{-i\phi}+e^{i\phi}e^{2i\omega_{mw}t} & \frac{\omega_{mw} - \omega_0}{\omega_R}\\
\end{bmatrix}.
\end{eqnarray}
By neglecting the high frequency oscillations ($2\omega_{mw}t$), the time-independent Hamiltonian in the rotating wave approximation is obtained:
\begin{eqnarray}
&&H = \frac{1}{2}\cdot\begin{bmatrix}
 \omega_{mw} - \omega_0 & \omega_R\cdot e^{i\phi}\\
 \omega_R\cdot e^{-i\phi} & \omega_{mw} - \omega_0\\
\end{bmatrix}
\\            
\label{eq:app_ham_rwa}
&=& (\omega_{mw} - \omega_0) \frac{\sigma_z}{2} + \omega_{R} \left[ \cos(\phi) \frac{\sigma_x}{2} - \sin(\phi) \frac{\sigma_y}{2} \right].
\end{eqnarray}

The ideal unitary operation is obtained by evaluating $U = e^{-i \cdot H \cdot T}$ for an ideal microwave signal ($\omega_{mw}$ = $\omega_0$):
\begin{eqnarray}
&&U_{ideal} = e^{-i \cdot \theta \cdot \left[ \cos(\phi) \cdot \frac{\sigma_x}{2} - \sin(\phi) \cdot \frac{\sigma_y}{2} \right]}
\\            
&=& \cos\left(\frac{\theta}{2}\right) I - i \sin\left(\frac{\theta}{2}\right)\left[\cos(\phi) \sigma_x - \sin(\phi) \sigma_y\right],
\end{eqnarray}
where $I$ is the identity matrix, and $\theta = \omega_R \cdot T$ the rotation angle and $\phi$ the rotation axis. This unitary is used to evaluate the process fidelity (Eq. \ref{eq:app_fidelity_definition} with $n = 2$) of any non-ideal operation $U_{real}$ attempting to implement the same rotation angle around the same rotation axis.

\subsection{Inaccuracies}
\label{A:1qubit_inacc}
Due to inaccuracies in the Hamiltonian or in the timing, the implemented operation $U_{real} = e^{-i \cdot H_{real} \cdot T_{real}}$ has a reduced fidelity.

In case of a microwave frequency inaccuracy $\omega_{mw} = \omega_{0} + \Delta\omega_{mw}$, the fidelity follows (for any rotation angle/axis) as:
%
\begin{equation}
\label{eq:app_acc_mw}
F = \frac{\left[\sin(\frac{\theta}{2})\sin(\frac{\theta}{2}\sqrt{\alpha^2+1}) + \sqrt{\alpha^2+1} \cos(\frac{\theta}{2})\cos(\frac{\theta}{2}\sqrt{\alpha^2+1})\right]^2}{\alpha^2 + 1},
\end{equation}
%
where $\alpha = \Delta\omega_{mw}/\omega_R$ is the error relative to the Rabi frequency. Taking the Taylor series expansion of Eq. \ref{eq:app_acc_mw} with respect to $\alpha$ leads to:
\begin{equation}
\label{eq:app_acc_mw_series}
F \approx 1 - \frac{1 - \cos(\theta)}{2} \cdot \alpha^2 + \mathcal{O}(\alpha^4).
\end{equation}

In case of a microwave phase inaccuracy $\phi = \phi_{ideal} + \Delta\phi$, the fidelity follows (for any rotation angle/axis) as:
\begin{equation}
\label{eq:app_acc_phi}
F = \left[\cos(\alpha)\sin^2\left(\frac{\theta}{2}\right)+\cos^2\left(\frac{\theta}{2}\right)\right]^2,
\end{equation}
where $\alpha = \Delta\phi$ is the error. Taking the Taylor series expansion of Eq. \ref{eq:app_acc_phi} with respect to $\alpha$ again leads to:
\begin{equation}
\label{eq:app_acc_phi_series}
F \approx 1 - \frac{1 - \cos(\theta)}{2} \cdot \alpha^2 + \mathcal{O}(\alpha^4).
\end{equation}

In case of an inaccuracy in the microwave amplitude $\omega_R = \omega_{R,ideal} + \Delta\omega_R$, the fidelity follows (for any rotation angle/axis) as:
\begin{equation}
\label{eq:app_acc_amp}
F = \cos^2\left(\frac{\theta}{2} \cdot \alpha \right),
\end{equation}
where $\alpha = \Delta\omega_{R}/\omega_{R,ideal}$ is the relative error. Taking the Taylor series expansion of Eq. \ref{eq:app_acc_amp} with respect to $\alpha$ leads to:
\begin{equation}
\label{eq:app_acc_amp_series}
F \approx 1 - \left(\frac{\theta}{2}\right)^2 \cdot \alpha^2 + \mathcal{O}(\alpha^4).
\end{equation}

In case of an inaccuracy in the microwave duration $T = T_{ideal} + \Delta T$, the fidelity follows (for any rotation angle/axis) again as:
\begin{equation}
\label{eq:app_acc_time}
F = \cos^2\left(\frac{\theta}{2} \cdot \alpha \right),
\end{equation}
where $\alpha = \Delta T/T_{ideal}$ is the relative error. Taking the Taylor series expansion of Eq. \ref{eq:app_acc_time} with respect to $\alpha$ will lead to Eq. \ref{eq:app_acc_amp_series}.\\
\\
A Z-rotation can be obtained without applying a signal to the qubit, simply by updating the reference frame, i.e. the phase of the microwave oscillator. The operation is given by:
\begin{equation}
    U = \begin{bmatrix}
    e^{i \phi / 2} & 0 \\
    0 & e^{-i \phi / 2} \\
    \end{bmatrix},
\end{equation}
where for the ideal operation $U_{ideal}$, the rotation angle $\phi = \phi_{ideal}$, and for the implemented operation $U_{real}$, the rotation angle $\phi = \phi_{ideal} + \Delta\phi$ has an error $\Delta\phi$. The fidelity follows (for any rotation angle) as:
\begin{equation}
\label{eq:app_acc_zrot}
F = \cos^2\left(\frac{\Delta\phi}{2} \right).
\end{equation}
Taking the Taylor series expansion of Eq. \ref{eq:app_acc_zrot} with respect to $\Delta\phi$ leads to:
\begin{equation}
\label{eq:app_acc_zrot_series}
F \approx 1 - \frac{1}{4} \cdot \Delta\phi^2 + \mathcal{O}(\Delta\phi^4).
\end{equation}

\subsection{Quasi-static Noise}
\label{A:1qubit_quasi}
The fidelity equations \ref{eq:app_acc_mw_series}, \ref{eq:app_acc_phi_series} and \ref{eq:app_acc_amp_series} all follow the relation $F(x) = 1 - c \cdot x^2$ for which the expected fidelity in case of Gaussian distributed inaccuracies (noise) is known (Eq. \ref{eq:app_gauss_2}). For all cases but the case of a microwave frequency inaccuracy, the expected fidelity can also be evaluated using the exact fidelity formula (no series expansion), which is more accurate for large amounts of noise.

In case of Gaussian distributed microwave phase noise $\phi = \mathcal{N}(\phi_{ideal}, \sigma_\phi^2)$, the expected fidelity follows (for any rotation angle/axis) as:
\begin{equation}
F = \frac{1}{2} \left(1+e^{-2\alpha^2}\right)\sin^4\left(\frac{\theta}{2}\right)+\cos^4\left(\frac{\theta}{2}\right)+\frac{1}{2} e^{-\frac{\alpha^2}{2}}\sin^2(\theta),
\end{equation}
where $\alpha = \sigma_\phi$ is the standard deviation of $\phi$.

In case of Gaussian distributed microwave amplitude noise $\omega_R = \mathcal{N}(\omega_{R,ideal}, \sigma_{\omega_R}^2)$, the expected fidelity follows (for any rotation angle/axis) as:
\begin{equation}
F = \frac{1}{2}+\frac{1}{2} e^{-\frac{1}{2}\alpha^2\theta^2},
\end{equation}
where $\alpha = \sigma_{\omega_R}/\omega_{R,ideal}$ is the relative standard deviation of $\omega_R$.

In case of Gaussian distributed timing variations $T = \mathcal{N}(T_{ideal}, \sigma_T^2)$, the expected fidelity follows (for any rotation angle/axis) again as:
\begin{equation}
F = \frac{1}{2}+\frac{1}{2} e^{-\frac{1}{2}\alpha^2\theta^2},
\end{equation}
where $\alpha = \sigma_T/T_{ideal}$ is the relative standard deviation of $T$.

\subsection{Noise Filtering}
\label{A:1qubit_filt}
In \cite{green2012high,Green2013} it is shown that when writing the total Hamiltonian $H(t) = H_c(t) + H_0(t)$ as the sum of a noise-free Hamiltonian $H_c(t)$ and a generalized noise Hamiltonian $H_0(t) = \beta_x(t) \sigma_x + \beta_y(t) \sigma_y + \beta_z(t) \sigma_z$, in first order approximation the expected process fidelity (from here on simply denoted with $F$) can be written as:
\begin{equation}
\label{eq:app_fidelity_spectra}
F = 1 - \frac{1}{2\pi}\sum\limits_{i,j,k=x,y,z} \int_{-\infty}^{\infty} S_{ij}(\omega) \cdot \frac{R_{jk}(\omega) R_{ik}^{*}(\omega)}{\omega^2} d\omega,
\end{equation}
with $S_{ij}(\omega)$ the cross-power spectral density between the random variables $\beta_i(t)$ and $\beta_j(t)$. The factors $R_{ij}(\omega)$, i.e. the control matrices in the frequency domain, depend on the control propagator $U_c(t) = e^{-i H_c(t) t}$:
\begin{equation}
R_{ij}(\omega) = -\frac{i \omega}{2} \int_0^{T} \mathrm{Tr} \left[U_c^{\dag}(t) \sigma_i U_c(t) \sigma_j \right] \cdot e^{i \omega t} dt.
\end{equation}

For the single-qubit operation, Eq. \ref{eq:app_ham_rwa} is used to represent the noise-free Hamiltonian $H_c(t)$. The high-frequency noise sources of interest are fluctuations in the microwave frequency $\omega_{mw}(t) = \omega_{mw,nom} + \delta\omega_{mw}(t)$, and fluctuations in the microwave envelope $\omega_R(t) = \omega_{R,nom} + \delta\omega_R(t)$. It is safe to assume that the fluctuations $\delta\omega_{mw}(t)$ and $\delta\omega_R(t)$ (with power spectral densities $S_{mw}(\omega)$ and $S_{R}(\omega)$, respectively) are statistically independent due to the different nature of the noise source.

With these assumptions, the generalized noise Hamiltonian follows as:
\begin{equation}
H_0(t) = \delta\omega_{mw}(t) \cdot \frac{\sigma_z}{2} + \delta\omega_{R}(t) \cdot \left[ \cos(\phi) \cdot \frac{\sigma_x}{2} - \sin(\phi) \cdot \frac{\sigma_y}{2} \right],
\end{equation}
and the non-zero cross-power spectral densities $S_{ij}(\omega)$ are found as:
\begin{eqnarray}
S_{xy}(\omega) &=& -\frac{1}{4} S_{R}(\omega) \sin(\phi)\cos(\phi)\\
S_{yx}(\omega) &=& -\frac{1}{4} S_{R}(\omega) \sin(\phi)\cos(\phi) \\
S_{xx}(\omega) &=& \frac{1}{4} S_{R}(\omega) \cos(\phi)^2\\
S_{yy}(\omega) &=& \frac{1}{4} S_{R}(\omega) \sin(\phi)^2 \\
S_{zz}(\omega) &=& \frac{1}{4} S_{mw}(\omega).
\end{eqnarray}

Eq. \ref{eq:app_fidelity_spectra} can now be evaluated to find the expected fidelity in case of amplitude noise (Eq. \ref{eq_filt_R}) and microwave frequency noise (Eq. \ref{eq_filt_mw}).
\begin{eqnarray}
\label{eq_filt_R}
F &=& 1 - \frac{1}{2\pi} \int_{-\infty}^{\infty} \frac{S_{R}(\omega)}{\omega_R^2} \cdot \left| H_{R}(\omega) \right|^2  \cdot d\omega
\\
\label{eq_filt_mw}
F &=& 1 - \frac{1}{2\pi} \int_{-\infty}^{\infty} \frac{S_{mw}(\omega)}{\omega_R^2}
\cdot \left| H_{mw}(\omega) \right|^2 \cdot d\omega,
\end{eqnarray}
where $\left| H_{R}(\omega) \right|^2$ and $\left| H_{mw}(\omega) \right|^2$ are the amplitude responses of the filter functions for the respective type of noise:
\begin{eqnarray}
\label{eq_filtH_R}
\left| H_{R}(\omega) \right|^2 &=& \frac{\sin\left( \alpha\frac{\theta}{2} \right)^2}{\alpha^2}
\\
\label{eq_filtH_mw}
\left| H_{mw}(\omega) \right|^2 &=& \frac{\left[1 - \cos(\theta) \cos(\alpha \theta) \right] \left(\alpha^2+1\right) -2 \alpha \sin(\theta)\sin(\alpha \theta)}{2\left(\alpha^2-1\right)^2},
\end{eqnarray}
where $\alpha = \frac{\omega} {\omega_R}$, is the frequency normalized to the Rabi frequency.
\\
\\
In case of wideband additive noise ($\delta\omega_{add}(t)$), the signal can be better modeled as $\omega_{ESR}(t) = 2 \left[ \omega_{R} \cdot \cos\left(\omega_{mw} t\right) + \delta\omega_{add}(t)\right]$. The lab frame Hamiltonian (Eq. \ref{eq:app_ham_lab}) follows as:
\begin{equation}
H_{lab} = -\omega_0 \cdot \frac{\sigma_z}{2} + \left[ \cdot \omega_{R} \cdot \cos\left(\omega_{mw} t\right) + \delta\omega_{add}(t)\right] \cdot \sigma_x.
\end{equation}
Evaluating this Hamiltonian in the rotating frame (Eqs. \ref{eq:app_rwa_1}-\ref{eq:app_rwa_3} with $\omega_{mw} = \omega_0$) leads to (after taking the RWA):
\begin{equation}
H = \omega_R \cdot \frac{\sigma_x}{2} + \delta\omega_{add}(t) \cdot \left[\cos(t \omega_0) \cdot \sigma_x + \sin(t \omega_0) \cdot \sigma_y \right].
\end{equation}
The first part of this equation is the noise-free Hamiltonian ($H_c(t) = \omega_{R} \cdot \frac{\sigma_x}{2}$), while the remainder forms the generalized noise Hamiltonian, with:
\begin{eqnarray}
\beta_x(t) &=& \delta\omega_{add}(t) \cdot \cos(t \omega_0)
\\
\beta_y(t) &=& \delta\omega_{add}(t) \cdot \sin(t \omega_0)
\\
\beta_z(t) &=& 0.
\end{eqnarray}
Assuming the additive noise $\delta\omega_{add}(t)$ has a power spectral density $S_{add}(\omega)$, the non-zero cross-power spectral densities $S_{ij}(\omega)$ are found as the Fourier transform of the cross-correlations:
\begin{equation}
R_{ij}(\tau) = \int_{-\infty}^{\infty} \beta_i(t) \beta_j(t + \tau) dt,
\end{equation}
which evaluate to:
\begin{eqnarray}
R_{xy}(\tau) &=& \frac{1}{2} \sin(\omega_0 \tau) \cdot R_{add}(\tau)
\\
R_{yx}(\tau) &=& -\frac{1}{2} \sin(\omega_0 \tau) \cdot R_{add}(\tau)
\\
R_{xx}(\tau) &=& \frac{1}{2} \cos(\omega_0 \tau) \cdot R_{add}(\tau)
\\
R_{yy}(\tau) &=& \frac{1}{2} \cos(\omega_0 \tau) \cdot R_{add}(\tau),
\end{eqnarray}
where $R_{add}(\tau)$ is the auto-correlation of $\delta\omega_{add}(t)$, i.e. the Fourier transform of $S_{add}(\omega)$. The $\sin(\omega_0 \tau)$ or $\cos(\omega_0 \tau)$ modulates the spectrum $S_{add}(\omega)$, leading to:
\begin{eqnarray}
S_{xy}(\omega) &=& \frac{1}{4} \left[ S_{add}(\omega + \omega_0) - S_{add}(\omega - \omega_0) \right]
\\
S_{yx}(\omega) &=& \frac{1}{4} \left[ S_{add}(\omega - \omega_0) - S_{add}(\omega + \omega_0) \right]
\\
S_{xx}(\omega) &=& \frac{1}{4} \left[ S_{add}(\omega + \omega_0) + S_{add}(\omega - \omega_0) \right]
\\
S_{yy}(\omega) &=& \frac{1}{4} \left[ S_{add}(\omega + \omega_0) + S_{add}(\omega - \omega_0) \right].
\end{eqnarray}
And the expected fidelity evaluates to (using the symmetry of the power spectral density):
\begin{equation}
F = 1 - \frac{1}{\pi} \int_{0}^{\infty} \frac{S_{add}(\omega - \omega_0)}{\omega_R^2} \cdot \left| H_{add}(\omega) \right|^2  \cdot d\omega,
\end{equation}
with:
\begin{equation}
\left| H_{add}(\omega) \right|^2 = \left| H_{R}(\omega) \right|^2 + \left| H_{mw}(\omega) \right|^2.
\end{equation}
\\

\subsection{Frequency Multiplexing}
\label{A:1qubit_fdma}
Driving a certain qubit at a frequency $\omega_0 = \omega_{mw}$ can also influences another qubit at a frequency $\omega_{0,other} = \omega_0 + \omega_{0,space}$ separated by $\omega_{0,space}$. To simplify the analysis, again a time-independent Hamiltonian is obtained by moving to a frame rotating with $\omega_{mw}$ (Eqs. \ref{eq:app_rwa_1}-\ref{eq:app_rwa_3}). In this frame, the other qubit appears to rotate around the z-axis with a frequency $\omega_{0,space}$. The ideal operation, an identity, can be described in this frame as:
\begin{equation}
U_{ideal} = e^{-i \cdot \omega_{0,space} \cdot \frac{\sigma_z}{2} \cdot T},
\end{equation}
whereas the real operation follows as (for $\phi = 0$):
\begin{equation}
\label{eq:app_fdma_real1}
U_{real} = e^{-i \cdot \left[\omega_{0,space} \cdot \frac{\sigma_z}{2} + \omega_{R,other} \cdot \frac{\sigma_x}{2}\right] \cdot T}.
\end{equation}
Recall that the Rabi frequency $\omega_{R,other}$ is related to the amplitude of the driving magnetic field by $\omega_{R,other} = A_{other} \cdot \gamma_e / 2$. In general, the required microwave amplitude for a certain Rabi frequency can be different for the two qubits involved. Therefore $\omega_{R,other}$ in the equation above can be considered as the amplitude driving the other qubit, while $\omega_R$ is introduced as the amplitude driving a rotation $\theta = \omega_R \cdot T$ in a duration $T$ on the intended qubit. Equation \ref{eq:app_fdma_real1} can then be rewritten as:
\begin{equation}
U_{real} = e^{-i \cdot \theta \cdot \left[\frac{\omega_{0,space}}{\omega_R} \cdot \frac{\sigma_z}{2} + \frac{\omega_{R,other}}{\omega_R} \cdot \frac{\sigma_x}{2}\right]}.
\end{equation}

The fidelity follows as ($\alpha = \frac{\omega_{0,space}}{\omega_R}$ and $\beta = \frac{\omega_{R,other}}{\omega_R}$):
\begin{widetext}
\begin{eqnarray}
F &=& \frac{\left|\sqrt{\alpha^2 + \beta^2}\cdot\cos\left(\frac{\theta}{2}\sqrt{\alpha^2 + \beta^2}\right)\cdot\left(1+\cos(\theta\alpha)+i\sin(\theta\alpha)\right)+\alpha\cdot\sin\left(\frac{\theta}{2}\sqrt{\alpha^2 + \beta^2}\right)\cdot\left(i-i\cos(\theta\alpha)+\sin(\theta\alpha)\right)\right|^2}{4\left(\alpha^2 + \beta^2\right)}.
\end{eqnarray}
\end{widetext}

Recall from the fidelity formula (Eq. \ref{eq:app_fidelity_definition}) that the fidelity is unity in case $U_{ideal}^\dag U_{real} = I$ ($I$ is the Identity). To gain more insight into the infidelity of the other qubit, the decomposition $U_{ideal}^\dag U_{real} = x\cdot\sigma_x+y\cdot\sigma_y+z\cdot\sigma_z+l\cdot I$ is made. Since $|l|^2 = F$ and $|x|^2 + |y|^2 + |z|^2 + |l|^2 = 1$, the infidelity equals $1-F = |x|^2 + |y|^2 + |z|^2$ and has contributions from X, Y and Z-rotations. Since the Z-rotations can easily be removed by a software update of the reference frame, the residual infidelity contributions ($1-F_{corr}$) can be found from the decomposition as:
\begin{equation}
|x|^2 + |y|^2 = \frac{\beta^2}{\alpha^2+\beta^2}\cdot\sin^2\left(\frac{\theta}{2}\sqrt{\alpha^2+\beta^2}\right).
\end{equation}
For $\alpha^2 \gg \beta^2$ (valid for the cases of interest: sufficient qubit spacing and $\omega_{R,other} \le \omega_R$):
\begin{equation}
F_{corr} \approx 1-\frac{\beta^2}{\alpha^2}\cdot\sin^2\left(\frac{\theta}{2}\alpha\right) \ge 1-\frac{\beta^2}{\alpha^2},
\end{equation}
which shows a close resemblance to the power of the Fourier transform of the rectangular microwave envelope: $4/\alpha^2\cdot\sin^2\left(\frac{\theta}{2}\alpha\right)$. 

\subsection{Idle Gate}
\label{A:1qubit_nop}
In case no microwave signal is applied to the qubits, residual noise on the drive line can still affect the qubits. Consider the Hamiltonian in the lab frame (Eq. \ref{eq:app_ham_lab}) with $\omega_{ESR}(t) = 2\cdot\omega_{R_n}(t)$, where $\omega_{R_n}(t)$ is the noise signal with spectral density $S_{R_n}(\omega)$. This Hamiltonian can again be split into a noise-free Hamiltonian ($H_c = -\omega_0\cdot\sigma_z/2$) and a generalized noise Hamiltonian $H_0(t) = \omega_{R_n}(t)\cdot\sigma_x$. It follows that $S_{xx}(\omega) = S_{R_n}(\omega)$ and all other cross-spectral densities are zero.

Eq. \ref{eq:app_fidelity_spectra} can now be evaluated, leading to:
\begin{eqnarray}
F &=& 1 - \frac{1}{\pi} \int_{-\infty}^{\infty} S_{R_n}(\omega) \cdot G(\omega) \cdot d\omega
\\
G(\omega) &=& \frac{\sin^2\left[\frac{T(\omega+\omega_0)}{2}\right]}{(\omega+\omega_0)^2}+\frac{\sin^2\left[\frac{T(\omega-\omega_0)}{2}\right]}{(\omega-\omega_0)^2}.
\end{eqnarray}
Using the symmetry of the spectrum $S_{R_n}(\omega)$ and by assuming $T = \theta / \omega_R$ is the time that would be needed to rotate a qubit by an angle $\theta$ when applying the signal amplitude for a Rabi frequency $\omega_R$:
\begin{equation}
F = 1 - \frac{1}{\pi} \int_{0}^{\infty} \frac{S_{R_n}(\omega)}{\omega_R^2}\cdot |H_n(\omega)|^2 \cdot d\omega,
\end{equation}
where
\begin{equation}
\label{eq:app_nop_filter}
|H_n(\omega)|^2 = 2\cdot\left(\frac{\omega_R}{\omega-\omega_0}\right)^2 \cdot \sin^2\left(\frac{\theta}{2}\frac{\omega-\omega_0}{\omega_R}\right),
\end{equation}
%
%
which represents the amplitude response of a sinc-shaped band-pass filter centered around $\omega_0$. For $\omega_0 = 0$ this would be a low-pass filter with effective noise bandwidth $ENBW_n = \omega_R\cdot \pi/|\theta|$ and DC-gain $|H_n(0)|^2 = \theta^2 / 2$. Therefore, a brickwall approximation of the amplitude response of Eq. \ref{eq:app_nop_filter}, which would be a good approximation in case of white noise, is:
\begin{equation}
|H_n(\omega)|^2 \approx
\begin{cases}
\theta^2 / 2 & |\omega-\omega_0| \le \omega_R\cdot \pi/|\theta|\\
0 &\text{elsewhere}
\end{cases}.
\end{equation}
\\
\\
Frequency inaccuracies also affect the qubits while no operation is performed. In a frame rotating with the oscillators frequency, the qubit appears to rotate with a frequency $\Delta\omega$ representing the frequency inaccuracy. The fidelity of an identity operation for a duration $T_{nop}$ evaluates to:
\begin{equation}
F = \cos^2\left(\frac{\Delta\omega \cdot T_{nop}}{2}\right),
\end{equation}
for which the Taylor series expansion follows as:
\begin{equation}
\label{eq:app_nop_w0}
F = 1 - \frac{T_{nop}^2}{4} \cdot \Delta\omega^2 + \mathcal{O}(\Delta\omega^4).
\end{equation}

A residual spurious tone driving the qubit for a duration $T_{nop}$ while not intended would also reduce its fidelity. Using the rotating frame Hamiltonian of Eq. \ref{eq:app_ham_rwa} with $\omega_0$ = $\omega_{mw}$ and $\omega_R$ = $\omega_{R,spur}$ results in a fidelity of an identity operation of:
\begin{equation}
F = \cos^2\left(\frac{\omega_{R,spur} \cdot T_{nop}}{2}\right),
\end{equation}
for which the Taylor series expansion follows as:
\begin{equation}
F = 1 - \frac{T_{nop}^2}{4} \cdot \omega_{R,spur}^2 + \mathcal{O}(\omega_{R,spur}^4).
\end{equation}

\section{Derivations for Two-Qubit Operation}
\label{A:2qubit}
As discussed in the main text, the following Hamiltonian is used for our 2-qubit system ($\hbar = 1$) \cite{coish2007exchange,meunier2011efficient,Veldhorst2015}:
\begin{equation}
\label{eq:app_2_ham}
H = \begin{bmatrix}
 -\omega_0  & 0 & 0 & 0 & 0 & 0 \\
 0 & \frac{\delta\omega_{0}}{2} & 0 & 0 & t_0 & t_0 \\
 0 & 0 & -\frac{\delta\omega_{0}}{2} & 0 & -t_0 & -t_0 \\
 0 & 0 & 0 & \omega_{0} & 0 & 0 \\
 0 & t_0 & -t_0 & 0 & U - \epsilon & 0 \\
 0 & t_0 & -t_0 & 0 & 0 & U + \epsilon \\
\end{bmatrix}.
\end{equation}

\subsection{Hamiltonian Eigenenergies}
\label{A:2qubit_lamba}
The Hamiltonian of Eq.\,\ref{eq:app_2_ham} has six eigenvalues. However, as the quantum state is encoded in the spin state, the two eigenvalues related to the single-dot singlet states are not analyzed in the following. The remaining four eigenvalues ($\omega_{\lambda_i}$) of the Hamiltonian (Eq.\,\ref{eq:app_2_ham}) have been analyzed for various Larmor frequency differences $\delta\omega_{0}$. However, independent of the choice of $\delta\omega_{0}$:
\begin{eqnarray}
    \omega_{\lambda_1} &=& -\omega_0 \\
    \omega_{\lambda_4} &=& \omega_0.
\end{eqnarray}

First, the case of $\delta\omega_{0} = 0$ is analyzed. By taking the 2$^{\text{nd}}$-order Taylor series expansion around $t_0 = 0$, i.e.~the tunnel coupling small compared to the charging energy, the eigenenergies are found as:
\begin{eqnarray}
    \label{eq:app_2_lambda_a_1}
    \omega_{\lambda_2} &=& 0 \\
    \label{eq:app_2_lambda_a_2}
    \omega_{\lambda_3} &=& - \frac{4 U t_0^2}{U^2 - \epsilon^2}.
\end{eqnarray}

Now, the special case of $\delta\omega_{0} = \sqrt{2} t_0$ is analyzed. By again taking the 2$^{\text{nd}}$-order Taylor series expansion around $t_0 = 0$, the eigenenergies are found as:
\begin{eqnarray}
    \omega_{\lambda_2} &=& \frac{\delta\omega_{0}}{2} - \frac{2 U t_0^2}{U^2 - \epsilon^2} \\
    \omega_{\lambda_3} &=& -\frac{\delta\omega_{0}}{2} - \frac{2 U t_0^2}{U^2 - \epsilon^2}.
\end{eqnarray}
When removing the Larmor precession:
\begin{eqnarray}
    \label{eq:app_2_lambda_b_1}
    \omega_{\lambda_2}' &=& - \frac{2 U t_0^2 }{U^2 - \epsilon^2} \\
    \label{eq:app_2_lambda_b_2}
    \omega_{\lambda_3}' &=& - \frac{2 U t_0^2 }{U^2 - \epsilon^2}.
\end{eqnarray}
Comparing Eqs. \ref{eq:app_2_lambda_a_1} and \ref{eq:app_2_lambda_a_2} with Eqs. \ref{eq:app_2_lambda_b_1} and \ref{eq:app_2_lambda_b_2}, it appears that $ \omega_{\lambda_2} +  \omega_{\lambda_3}$ is independent of the choice of $\delta\omega_{0}$, but that the fraction $\omega_{\lambda_2}'/\omega_{\lambda_3}'$ changes.

A more in depth analysis shows that the exact, and generally valid, solutions to $\omega_{\lambda_2}$ and $\omega_{\lambda_3}$ are found by solving Eq. \ref{eq:app_2_lambda_general}. Under the assumptions $\omega_{\lambda_i} \ll t_0, U$ and $U^2 - e^2 \gg t_0^2, \delta\omega_0^2$ and $\delta\omega_0^2 \ll 8 t_0^2$, this equation can be simplified to Eq. \ref{eq:app_2_lambda_simple}.

\begin{widetext}
\begin{eqnarray}
    \label{eq:app_2_lambda_general}
    \omega_{\lambda_i}^4 - 4 \omega_{\lambda_i}^3 U  + \omega_{\lambda_i}^2 \left(-\delta\omega_{0}^2 -4 \epsilon^2 -16 t_0^2 +4 U^2\right) + 4 \omega_{\lambda_i} U \left(\delta\omega_{0}^2 + 8 t_0^2 \right) + 4 \delta\omega_{0}^2 \left(\epsilon^2 - U^2\right) &=& 0 \\
    \label{eq:app_2_lambda_simple}
    \omega_{\lambda_i}^2 \left(-4 \epsilon^2 +4 U^2\right) + 4 \omega_{\lambda_i} U 8 t_0^2 + 4 \delta\omega_{0}^2 \left(\epsilon^2 - U^2\right) &\approx& 0
\end{eqnarray}
\end{widetext}

\begin{figure*}
\includegraphics[width=\linewidth]{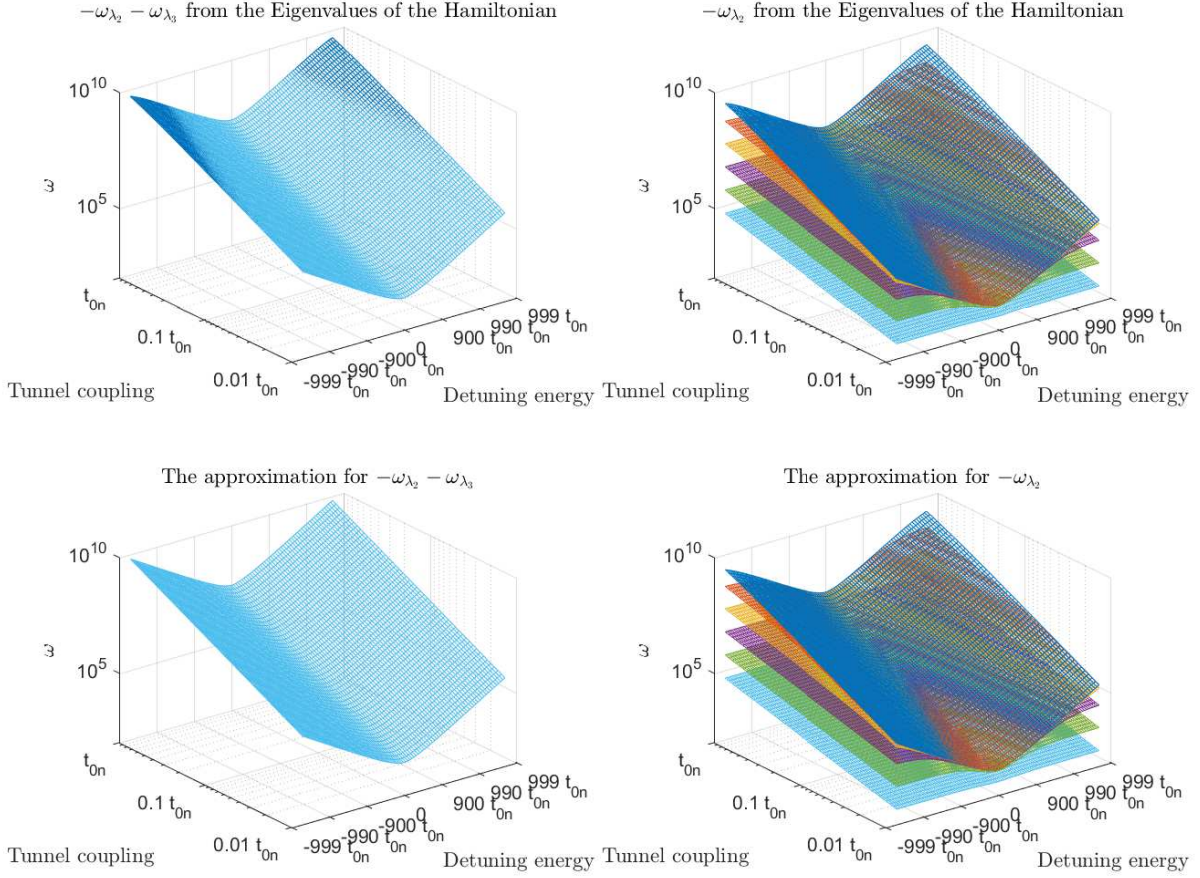}
\caption{\label{fig:app_2_speed} The 2-qubit operation speeds $\omega_{\lambda_2}$ and $\omega_{\lambda_3}$ versus the interdot tunnel coupling and detuning. A nominal tunnel coupling $t_{0n}$ of 1 GHz is used. The plots on the top show the eigenvalues of the Hamiltonian (Eq. \ref{eq:app_2_ham}), whereas the plots on the bottom show the approximation of Eqs. \ref{eq:app_2_lambda_b_1} and \ref{eq:app_2_lambda_b_2}. The different colors are used for different values of $\delta\omega_{0}$ ($\delta\omega_{0} = \sqrt{2} t_0 / 10^n$, with $n$ ranging from 0 for the blue curves upto 5 for the cyan curves.}
\end{figure*}

As a result, for the case $\delta\omega_{0} < \sqrt{2} t_0$, the eigenenergies are found as:
\begin{eqnarray}
    \label{eq:app_2_lambda_c_1}
    \omega_{\lambda_2} &=& - \frac{2 U t_0^2}{U^2 - \epsilon^2} +\frac{1}{2}\sqrt{\delta\omega_{0}^2 + \left( \frac{4 U t_0^2}{U^2 - \epsilon^2}\right)^2}\\
    \label{eq:app_2_lambda_c_2}
    \omega_{\lambda_3} &=& - \frac{2 U t_0^2}{U^2 - \epsilon^2} -\frac{1}{2}\sqrt{\delta\omega_{0}^2 + \left( \frac{4 U t_0^2}{U^2 - \epsilon^2}\right)^2}.
\end{eqnarray}
To show the validity of the simplification of Eq. \ref{eq:app_2_lambda_general} to Eq. \ref{eq:app_2_lambda_simple}, Fig.~\ref{fig:app_2_speed} compares the eigenvalues of the Hamiltonian (Eq. \ref{eq:app_2_ham}), with the eigenvalues as obtained from Eqs. \ref{eq:app_2_lambda_c_1} and \ref{eq:app_2_lambda_c_2}.

Hence, in general:
\begin{eqnarray}
\omega_{\lambda_1} &=& -\omega_0\\
\omega_{\lambda_2} &\approx& \begin{cases}\frac{-\omega_{op} + \sqrt{\delta\omega_{0}^2 + \omega_{op}^2}}{2}  & 0 \le \delta\omega_{0} < \sqrt{2} t_0 \\ \frac{-\omega_{op} + \delta\omega_{0}}{2}  & \delta\omega_{0} = \sqrt{2} t_0 \end{cases}\\
\omega_{\lambda_3} &\approx& \begin{cases} \frac{-\omega_{op} - \sqrt{\delta\omega_{0}^2 + \omega_{op}^2}}{2} & 0 \le \delta\omega_{0} < \sqrt{2} t_0 \\ \frac{-\omega_{op} - \delta\omega_{0}}{2} & \delta\omega_{0} = \sqrt{2} t_0 \end{cases}\\
\omega_{\lambda_4} &=& \omega_0,
\end{eqnarray}
where
\begin{equation}
\omega_{op} = 4 \cdot t_0^2 \cdot \frac{U}{U^2 - \epsilon^2}.
\end{equation}

\subsection{The C-Phase Gate}
\label{A:2qubit_cz}
To perform a C-phase gate, the control parameter must change adiabatically. An adiabatic change of the control implies that if the qubit state was an eigenvector (stationary state or eigenstate) of the Hamiltonian, it remains an eigenstate of the new Hamiltonian after the control change.
As a result, the ideal adiabatic operation to the desired operating point can be described as:
\begin{equation}
    U_{in} = V,
\end{equation}
where $V$ contains the eigenvectors of the Hamiltonian in the desired operating point $H_{op}$, assuming the Hamiltonian is expressed in a basis formed by the eigenvectors in the original operating point.

Moving back from the desired operating point to this original point can then be described as:
\begin{equation}
    U_{out} = V^{-1}.
\end{equation}

Finally, the operation in the desired operating point can be described as:
\begin{equation}
\label{eq:app_2_U_notadia}
    U_{op} = e^{-i H_{op} T} = V \cdot e^{-i D T} \cdot V^{-1},
\end{equation}
where the eigenvalue decomposition of $H_{op} = V \cdot D \cdot V^{-1}$ has been used ($D$ is a diagonal matrix containing the eigenvalues of $H_{op}$).

In total, an adiabatic operation ($U$) consisting of moving adiabatically to the desired operating point, operating for a while in this point, and moving back adiabatically, can be simplified as:
\begin{eqnarray}
    U &=& U_{out} \cdot U_{op} \cdot U_{in}
    \\
    &=& V^{-1} \cdot V \cdot e^{-i D T} \cdot V^{-1} \cdot V
    \\
    \label{eq:app_2_U_adia}
    &=& e^{-i D T},
\end{eqnarray}
which is then a diagonal matrix that only depends on the eigenenergies of the Hamiltonian:
\begin{equation}
\label{eq:app_2_u_cphase_lab}
U_{cz,lab}(t) = \begin{bmatrix}
 e^{-i  t  \omega_{\lambda_1}} & 0 & 0 & 0 \\
 0 & e^{-i  t  \omega_{\lambda_2}} & 0 & 0 \\
 0 & 0 & e^{-i  t  \omega_{\lambda_3}} & 0 \\
 0 & 0 & 0 & e^{-i  t  \omega_{\lambda_4}} \\
\end{bmatrix}.
\end{equation}
In the rotating frame this becomes the unitary operation describing the C-phase gate (Eq. \ref{eq:2_u_cphase}).

In case of an inaccuracy in the control parameters, the operation is still described by the diagonal matrix of Eq. \ref{eq:app_2_u_cphase_lab}, and takes the form of Eq. \ref{eq:2_u_cphase} in the rotating frame, however with different angles $\phi_{Z,A}$ and $\phi_{Z,B}$. The fidelity of an inaccurate operation follows as (Eq. \ref{eq:meth_static}):
\begin{equation}
\begin{split}
F &= \frac{3}{8} + \cos\left(\phi_{Z,A,ideal}-\phi_{Z,A,real} - \phi_{Z,B,ideal}+\phi_{Z,B,real}\right)\\
& + \frac{2}{8} \cos\left(\phi_{Z,A,ideal}-\phi_{Z,A,real}\right)\\
& + \frac{2}{8} \cos\left(\phi_{Z,B,ideal}-\phi_{Z,B,real}\right),
\end{split}
\end{equation}
where $\phi_{Z,A,ideal}$ and $\phi_{Z,B,ideal}$ are the acquired phases in case of no inaccuracy, and $\phi_{Z,A,real}$ and $\phi_{Z,B,real}$ are the acquired phases in case of an inaccuracy in the control parameter.

Evaluating this formula for inaccuracies in duration ($T_{real} = T + \Delta T$), tunnel coupling ($t_{0_{real}} = t_0 + \Delta t_0$), and detuning ($\epsilon_{real} = \epsilon + \Delta \epsilon$), for the different operating points, lead to the infidely formulas as summarized in Table \ref{tab:2_fidelities}, when taking the 2\textsuperscript{nd}-order Taylor series expansion to the inaccuracy (except for the case $\epsilon$ = 0 for which a 4\textsuperscript{th}-order Taylor series expansion is used).

\subsection{The Exchange Gate}
\label{A:2qubit_j}
In case the control parameter changes diabatically, as required for the exchange gate, $U_{in}$ and $U_{out}$ approximate the identity matrix, and the resulting 2-qubit operation is described by Eq. \ref{eq:app_2_U_notadia}. The relevant eigenenergies (in $D$) are given in Section \ref{A:2qubit_lamba}.

For $\delta\omega_{0} = 0$ and small $t_0$ (taking the Taylor series expansion), the $4\times4$ relevant entries of the eigenvector matrix can be approximated as:
\begin{equation}
V^{-1} = \begin{bmatrix}
 1 & 0 & 0 & 0 \\
 0 & \frac{1}{2} & \frac{1}{2} & 0 \\
 0 & -\frac{t_0}{U + \epsilon} & \frac{t_0}{U + \epsilon} & 0 \\
 0 & 0 & 0 & 1 \\
\end{bmatrix}.
\end{equation}
Eq. \ref{eq:app_2_U_notadia} can now be evaluated, leading to:
\begin{equation}
U_{J'}(t) \approx \begin{bmatrix}
 e^{-i t \omega_{\lambda_1}} & 0 & 0 & 0 \\
 0 & \frac{e^{-i t \omega_{\lambda_{2}}}+e^{-i t \omega_{\lambda_{3}}}}{2} & \frac{e^{-i t \omega_{\lambda_{2}}}-e^{-i t \omega_{\lambda_{3}}}}{2} & 0\\
 0 & \frac{e^{-i t \omega_{\lambda_{2}}}-e^{-i  t  \omega_{\lambda_{3}}}}{2} & \frac{e^{-i  t \omega_{\lambda_{2}}}+e^{-i  t  \omega_{\lambda_{3}}}}{2} & 0\\
 0 & 0 & 0 & e^{-i  t  \omega_{\lambda_4}} \\
\end{bmatrix}.
\end{equation}
In the rotating frame this becomes the unitary operation describing the Exchange gate (Eq. \ref{eq:2_u_exchange}).

In case of an inaccuracy in the control parameters, the operation is still described by this unitary matrix, however with a different angle $\theta_J$. The fidelity of an inaccurate operation follows as (Eq. \ref{eq:meth_static}):
\begin{equation}
F = \frac{5}{8} + \frac{3}{8} \cos\left(\theta_{J,ideal}-\theta_{J,real} \right),
\end{equation}
where $\theta_{J,ideal}$ is the acquired rotation angle in case of no inaccuracy, and $\theta_{J,real}$ is the acquired rotation angle in case of an inaccuracy in the control parameter.

Evaluating this formula for inaccuracies in duration ($T_{real} = T + \Delta T$), tunnel coupling ($t_{0_{real}} = t_0 + \Delta t_0$), and detuning ($\epsilon_{real} = \epsilon + \Delta \epsilon$), lead to the infidely formulas as summarized in Table \ref{tab:2_fidelities}, when taking the 2\textsuperscript{nd}-order Taylor series expansion to the inaccuracy (except for the case $\epsilon$ = 0 for which a 4\textsuperscript{th}-order Taylor series expansion is used).

\subsection{Idle Gate}
\label{A:2qubit_nop}
Evaluating the fidelity (Eq. \ref{eq:meth_static}) of the 2-qubit operations (Eq. \ref{eq:2_u_cphase} and Eq. \ref{eq:2_u_exchange}) with respect to an identity operation as ideal operation ($U_{ideal} = I$), leads to
\begin{equation}
\label{eq:2_app_nop_cz}
\begin{split}
F_I &= \frac{3}{8} + \cos\left(\phi_{Z,A} - \phi_{Z,B}\right)\\
& + \frac{2}{8} \cos\left(\phi_{Z,A}\right) +\frac{2}{8} \cos\left(\phi_{Z,B}\right)
\end{split}
\end{equation}
and
\begin{equation}
\label{eq:2_app_nop_J}
   F_I = \frac{5}{8} + \frac{3}{8} \cos\left(\theta_{J} \right),
\end{equation}
for the C-phase gate and exchange gate respectively.

Assuming a total acquired phase $\theta_{cz} = -(\phi_{Z,A} + \phi_{Z,B})$, and using the value of $\phi_{Z,B}$ as summarized in Table \ref{tab:2_fidelities} for different values of $\delta\omega_{0}$, simplifies Eq. \ref{eq:2_app_nop_cz} to:
\begin{equation}
F_I =
\begin{cases}
1 - \frac{3}{16} \cdot \theta_{cz}^2 & \delta\omega_0 = 0\\
1 - \frac{7-4\sqrt{2}}{16} \cdot \theta_{cz}^2 & \delta\omega_0 = \omega_{op}\\
1 - \frac{1}{16} \cdot \theta_{cz}^2 & \delta\omega_0 = \sqrt{2} t_0
\end{cases},
\end{equation}
after taking the 2\textsuperscript{nd} order Taylor series expansion to $\theta_{cz}$.

Taking the 2\textsuperscript{nd} order Taylor series expansion to $\theta_{J}$ in Eq. \ref{eq:2_app_nop_J} leads to:
\begin{equation}
F_I =
1 - \frac{3}{16} \cdot \theta_{J}^2.
\end{equation}

\section{Qubit Read-out}
\label{A:readout}
%
%
For the fidelity, here it is assumed that the post-measurement qubit state is of interest as well. As a result, the fidelity can be formulated as:
\begin{equation}
\label{eq:sup_read_fidelity_full}
    F = P_{charge} \cdot \left[P_{sense} \cdot P_{detect} + (1-P_{sense}) \cdot (1-P_{detect})\right],
\end{equation}
where also a sensing error ($1-P_{sense}$) together with a detection error ($1-P_{detect}$) could lead to the correct outcome, assuming these probabilities are uncorrelated. However, when larger fidelities are targeted, and hence smaller errors can be tolerated, a good approximation to Eq. \ref{eq:sup_read_fidelity_full} is given by:
\begin{equation}
    F \approx P_{charge} \cdot P_{sense} \cdot P_{detect}.
\end{equation}

The contribution $P_{charge}$ can be found from the system Hamiltonian. The Hamiltonian of Eq. \ref{eq:2_ham} is extended with the lowest-energy triplet states (spaced $E_{ST}$ from the singlet energy level). For the Hamiltonian, only the charge states with one electron in each dot and two electrons in the right dot are considered, i.e. in the basis $\Psi = \left[ \left| \uparrow, \uparrow \right\rangle, \left| \uparrow, \downarrow \right\rangle, \left| \downarrow, \uparrow \right\rangle, \left| \downarrow, \downarrow \right\rangle, \left| 0,\uparrow\uparrow \right\rangle, \left| 0,\uparrow \downarrow \right\rangle, \left| 0,\downarrow \uparrow \right\rangle, \left| 0,\downarrow\downarrow \right\rangle \right]$:

\begin{widetext}
\begin{equation}
\label{eq:readout_ham}
H = \begin{bmatrix}
 -\omega_0  & 0 & 0 & 0 & \sqrt{2} t_0 & 0 & 0 & 0\\
 0 & \frac{\delta\omega_{0}}{2} & 0 & 0 & 0 & \sqrt{2} t_0 & 0 & 0 \\
 0 & 0 & -\frac{\delta\omega_{0}}{2} & 0 & 0 & 0 & \sqrt{2} t_0 & 0 \\
 0 & 0 & 0 & \omega_{0} & 0 & 0 & 0 & \sqrt{2} t_0 \\
 \sqrt{2} t_0 & 0 & 0 & 0 & U - \epsilon + E_{ST} - \omega_0 & 0 & 0 & 0 \\
 0 & \sqrt{2} t_0 & 0 & 0 & 0 & U - \epsilon + \frac{E_{ST}}{2} & \frac{E_{ST}}{2} & 0 \\
 0 & 0 & \sqrt{2} t_0 & 0 & 0 & \frac{E_{ST}}{2} & U - \epsilon + \frac{E_{ST}}{2} & 0 \\
  0 & 0 & 0 & \sqrt{2} t_0 & 0 & 0 & 0 & U - \epsilon + E_{ST} + \omega_0 \\
\end{bmatrix}.
\end{equation}
\end{widetext}

To estimate $P_{detect}$, we assume Gaussian distributed noise and a simple measurement discrimination by comparison with a threshold $I_t = I_s/2$ halfway 0 and $I_s$ (i.e.~the signal for the two charge configurations to distinguish). For that case:
\begin{equation}
    P_{detect} = \frac{1}{2} + \frac{1}{2} \cdot \text{erf} \left( \frac{I_t}{\sigma_i \sqrt{2}} \right).
\end{equation}
For an integration time $T_{read}$, the filter transfer function is given by:
\begin{equation}
    H_{read}(\omega) = e^{-\frac{1}{2}i \omega T_{read}} \cdot \frac{2}{\omega} \cdot \sin\left(\frac{1}{2} \omega T_{read}\right).
\end{equation}
This filter function has an effective noise bandwidth $ENBW = 1/(2 T_{read})$. The standard deviation of the noise $\sigma_i$ can be approximated by integrating the noise power spectral density $S_i(f)$ (assumed flat) in the effective noise bandwidth (ENBW) of the filter. The standard deviation of the noise follows as $\sigma_i^2 = S_i / 2 / T_{read}$, leading to:
\begin{equation}
    P_{detect} = \frac{1}{2} + \frac{1}{2} \cdot \text{erf} \left( \frac{I_s/2}{\sqrt{S_i / T_{read}}} \right).
\end{equation}
%

\bibliography{paper}

\end{document}